\documentclass[aps,prl,twocolumn,superscriptaddress,showpacs,floatfix]{revtex4-2}

\usepackage{amsmath,amsthm,amssymb,amsfonts,float,graphics,epsfig,epstopdf,color,verbatim,tabularx,bm,multirow,appendix}
\usepackage[utf8]{inputenc}
\usepackage[T1]{fontenc}
\usepackage{xcolor}
\usepackage{dsfont}
\usepackage{textcomp}
\usepackage{yfonts}
\usepackage{footnote}
\usepackage{bm}
\usepackage{subfigure}
\usepackage{amsmath}
\usepackage{mathrsfs}
\usepackage{MnSymbol}

\newtheorem{criterion}{Criterion}
\usepackage{mathtools}
\usepackage{graphicx}
\usepackage{verbatim}
\usepackage[colorlinks=true, citecolor=blue, linkcolor=blue, urlcolor=blue]{hyperref}
\usepackage{multirow}
\usepackage{braket}
\usepackage[normalem]{ulem}
\usepackage{tikz}
\usetikzlibrary{calc}
\usetikzlibrary{shapes.multipart}
\usepackage{orcidlink}
\usepackage{xr-hyper} 
\usepackage[percent]{overpic} 

\newcommand{\hexagon}[2][0.4]{%
\begin{tikzpicture}[scale=#1,baseline=-0.5ex]
  \def\radius{1} 
  \foreach \i in {1,...,6} {
    \coordinate (P\i) at (-60*\i:\radius);
  }
  \draw (P1)--(P2)--(P3)--(P4)--(P5)--(P6)--cycle;
  \foreach \i in {1,...,6} {
    \node at (-60*\i:\radius+0.3) {#2\i};
  }
\end{tikzpicture}
}

\DeclareMathOperator{\Tr}{Tr}

\newcommand{\comments}[1]{}

\newcommand{\certGNME}{(\checkmark)}      
\newcommand{\mE}{\mathcal{E}}

\newcommand{\tabnotetext}[2]{%
  \par\raggedright{\footnotesize $^{#1}$ #2}%
}


\begin{document}

\title{Network-Irreducible Multiparty Entanglement in Quantum Matter}

\author{Liuke Lyu}
\affiliation{D\'epartement de Physique, Universit\'e de Montr\'eal, Montr\'eal, QC H3C 3J7, Canada}
\affiliation{
 Institut Courtois, Universit\'e de Montr\'eal, Montr\'eal (Qu\'ebec), H2V 0B3, Canada
}
\affiliation{
 Centre de Recherches Math\'ematiques, Universit\'e de Montr\'eal, Montr\'eal, QC, Canada, HC3 3J7
}
\author{Pedro Lauand}
\affiliation{Perimeter Institute for Theoretical Physics, Waterloo, Ontario, N2L 2Y5, Canada}
\affiliation{Institute for Quantum Computing, University of Waterloo, Waterloo, Ontario, N2L 3G1, Canada}

 
\author{William Witczak-Krempa}
\email{w.witczak-krempa@umontreal.ca}
\affiliation{D\'epartement de Physique, Universit\'e de Montr\'eal, Montr\'eal, QC H3C 3J7, Canada}
\affiliation{
 Institut Courtois, Universit\'e de Montr\'eal, Montr\'eal (Qu\'ebec), H2V 0B3, Canada
}
\affiliation{
 Centre de Recherches Math\'ematiques, Universit\'e de Montr\'eal, Montr\'eal, QC, Canada, HC3 3J7
}

\begin{abstract}
We show that the standard approach to characterize collective entanglement via genuine multiparty entanglement (GME) leads to an area law in ground and thermal Gibbs states of local Hamiltonians. To capture the truly collective part one needs to go beyond this short-range contribution tied to interfaces between subregions. Genuine network multiparty entanglement (GNME) achieves a systematic resolution of this goal by analyzing whether a $k$-party state can be prepared by a quantum network consisting of $(k-1)$-partite resources. We develop tools to certify and quantify GNME, and benchmark them for GHZ, W and Dicke states. We then study the 1d transverse field Ising model, where we find a sharp peak of GNME near the critical phase transition, and rapid suppression elsewhere. Finite temperature leads to a faster death of GNME compared to GME. Furthermore, certain 2d quantum spin liquids do not have GNME in microscopic subregions while possessing strong GME. 
 This approach will allow us to chart truly collective entanglement in quantum matter both in and out of equilibrium.  
\end{abstract}


\date{\today}
\maketitle

\begin{figure*}[hbt!]
    \centering
    \includegraphics[width=0.85\linewidth]{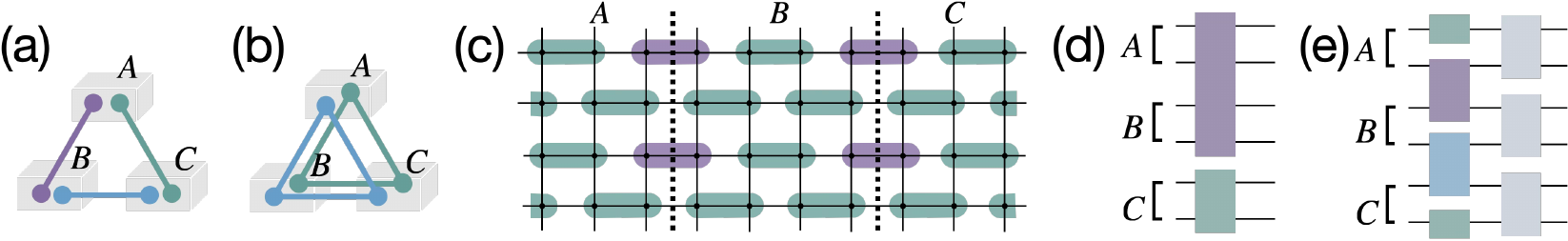}
    \caption{\textbf{Area-law of genuine multipartite entanglement and beyond.} 
    (a) Triangle network state consisting of three Bell pairs.
    (b) Two copies of the three-qubit GHZ state, which has network irreducible entanglement (GNME).
    (c) Valence bond solid (VBS) state.
    (d) Depth-1 circuit that prepares a biseparable state. 
    (e) Depth-2 brickwork circuit ($B_2$) that prepares a triangle network state.}
    \label{fig:network}
\end{figure*}

\paragraph{Introduction.} Quantum many-body systems, such as the ground and thermal Gibbs states of local Hamiltonians, possess a rich entanglement structure. A challenging question is to unravel the multiparty entanglement they encode, and its fate under various perturbations. A key concept has guided this quest: genuine multiparty entanglement (GME), which proves to be a sensitive probe for quantum criticality and exotic states of matter, whether applied to the global ground state~\cite{Hauke2016, Pezze2017} or to local subregions (entanglement microscopy)~\cite{Hofmann2014Scaling, Giampaolo2013XY, Giampaolo_2014,parez2026fate, Wang2025entanglement, Lyu2025multiparty, lyu2025loop, garcia2025fate, sabharwal2025characterizing}.
A $k$-party state has GME if it is not a mixture of states that are separable under some bipartition of the $k$ parties; such mixtures define biseparable states. A GME state is often said to possess collective entanglement, but is it ``truly'' collective? 
Consider a 3-party state $ABC$ where each party has a pair of qubits, and these form Bell states between the parties: one qubit of $A$ is maximally entangled with one qubit of Bob, and so forth, as illustrated in Fig.~\ref{fig:network}a. This pure state, which is an example of a Valence Bond Solid (VBS), has strong GME, but its multiparty entanglement is of bipartite origin. The same holds true for the VBS on a 2d lattice shown in Fig.~\ref{fig:network}c, where we further see the local nature of the GME. How can we distinguish such ``local'' GME from the more collective form expected in highly non-trivial states, such as the ground states at quantum critical phase transitions?

In this Letter, we show that the GME of ground or Gibbs states of local Hamiltonians is typically dominated by an area (boundary) law contribution. 
First, we quantify the area law in the case of a family of GME monotones that includes the genuine multiparty negativity (GMN)~\cite{Guhne2011,Hofmann2014}. 
This constitutes a limitation of GME as key universal properties of a state/phase are typically encoded in terms that are subleading to the area law, and one needs to devise ways to extract them, such as the topological entropy built from a linear combination of entropies~\cite{Eisert2010AreaLaw}.
As these subtraction methods come with fundamental limitations (such as the lack of an operational meaning), we argue that a systematic way to extract the ``truly collective'' multiparty entanglement lies in a more recent notion of GME that arose in quantum information and quantum foundations: Genuine Network Multiparty Entanglement (GNME)~\cite{navascues2020genuine,kraft2021quantum,luo2021new}. 
It applies to general mixed states and for an arbitrary number of parties. In the tripartite case, one asks whether a state can be prepared by sharing bipartite resources between the parties. Viewing each party as a node, this represents a triangle quantum network with bipartite connections. Although the above VBS states can be prepared with such a network, it is not possible to do so for all states. 
An example is a GHZ/cat state $\vert000\rangle+\vert111\rangle$,
or multiple copies thereof (Fig.~\ref{fig:network}b).
We argue that contrary to the GME, the GNME in typical ground or Gibbs states has sub-area-law scaling. 
We then develop tools to quantify the GNME. First, we consider the so-called inflation protocols~\cite{navascues2020genuine} that leverage semi-definite programming (SDP) to certify that a state has GNME. This approach yields strong results for a small number of qubits, but becomes more challenging as the Hilbert space dimension grows. 
Second, we develop optimization methods that allow us to obtain a strong upper bound to the distance between a given state and the convex set of network states. 

We benchmark our methods on tripartite canonical states of 3 and 6 qubits as well as 4-partite states of 8 qubits, yielding strong bounds on the white-noise robustness of GHZ, W, and Dicke states. A simple, yet non-trivial result is that the 3-qubit W-state $\ket{001}+\ket{010}+\ket{100}$ mixed 50\% white noise has GME but no GNME. 
Next, we investigate ground and Gibbs states of various physically relevant Hamiltonians in 1d/2d. We find that GNME is strongly suppressed compared to the GME, in agreement with the sub-area-law scaling of the former. In the 1d transverse field Ising model (TFIM), we find 1) a sharp peak of GNME near the quantum phase transition when considering adjacent parties; 2) GNME vanishes at a manifestly lower temperature than GME. In 2d, we discover that microscopic subregions of certain quantum spin liquids, including the paradigmatic Kitaev honeycomb model~\cite{Kitaev2006}, possess no GNME while having non-negligible GME. 

\paragraph{GME Area Law.} 
The VBS examples illustrate that GME is often dominated by bipartite entanglement across the interfaces between parties, leading to an area law in the limit of large subregions.
Generic ground and Gibbs states of local Hamiltonians are known to obey an area law for bipartite entanglement (logarithmic enhancements can occur in 1d critical systems, or fermionic states with a finite Fermi surface)~\cite{Eisert2010AreaLaw}. This will naturally lead to an area law for GME (and logarithmic enhancements for the same states), as we now explain.
Let us consider a bipartite entanglement monotone $E$. We can lift it to a GME measure via the following construction: in a $k$-party system, we first consider $E_{\min}(\rho) = \min_X E(\rho; X)$, where $X$ denotes a bipartition such as $1|2\cdots k$. We then take a mixed convex roof to get a GME measure:
\begin{equation}
    \mE = \min_{\sum_i p_i \rho_i}  \sum_i p_i E_{\min}(\rho_i) \; \leq \; E_{\min}(\rho)
    \label{eq:CRE_GME}
\end{equation}
where the upper bound immediately follows from the mixed roof. $ \mE $ is a GME monotone,  see SM Sec.\ref{sup:mixedCRE} for a proof. 
The upper bound means that the leading contribution (in the limit of large subregions) of $\mE$ is bounded by an area-law-dominated term, where the area corresponds to the smallest inter-party interface. We now argue that this is precisely the leading term of $\mE$. Let us begin with a ground state, and let the parties span the entire system so that the state for which we want to evaluate $\mE$ is pure. This leads to $\mE = E_{\min}(X^*)$, yielding the desired result:
\begin{equation}
\mE = \alpha_0 \times {\rm Area}(\;\text{min interface} \;) + \cdots
\label{eq:area_gme}
\end{equation}
where $\alpha_0$ is a non-negative coefficient, and the ellipsis denotes subleading corrections.
Here, we have used an $E$ that scales linearly with the area, such as the logarithmic negativity $LN$~\cite{plenio2005LN}. Recall that $LN=\log_2(2N+1)$, where $N$ is the negativity defined by the partial transpose with respect to one of the two parties~\cite{Peres1996}.
Now, if the parties do not span the entire system, the reduced density matrix becomes mixed. However, due to the locality of the Hamiltonian, the partial trace only induces high mixedness near the boundary between the k-party region A and its complement~\cite{Giudici2018, Dalmonte2022}. The bulk of A is thus nearly pure, and so we recover the leading result $\mE\approx E_{\min}(\rho)+...$, leading to an area law with the same coefficient as above. 

In a Gibbs state, the state of the entire system is given by the density matrix $\exp(-\beta H)/Z$, where $\beta=1/T$ is the inverse temperature (in units of the Boltzmann constant). At sufficiently small temperatures, continuity gives an area law with a modified coefficient $\alpha_0\to\alpha_T$. Furthermore, locality allows us to compute the coefficient by considering the bipartite problem of a single interface. We thus get that $\alpha_T$ is the area-law coefficient obtained by evaluating the bipartite measure $E$.

\paragraph{Truly Collective Entanglement via Networks.}
In order to understand the truly collective multiparty entanglement of a state, we change our perspective from GME to its network version, GNME~\cite{navascues2020genuine,kraft2021quantum,luo2021new}. 
Let us first review its definition in the tripartite case. To get a tripartite network state $\rho_{\rm net}$, we first distribute independent bipartite resource states along the three links of the network: $A_0B_0$, $B_1C_0$, $C_1A_1$, where each node is partitioned into two subsystems, $A=A_0A_1$ and similarly for $B$ and $C$. We then apply local completely positive trace-preserving (CPTP) maps at the nodes ($A$,$B$ and $C$), possibly chosen according to shared classical randomness with a probability distribution $p_\lambda$: 
\begin{equation}
    \rho_{\rm net} = \sum_\lambda p_\lambda C_\lambda(\sigma_{A_0B_0} \otimes \sigma_{B_1C_0} \otimes \sigma_{C_1A_1})
    \label{eq:network}
\end{equation}
where (omitting $\lambda$) $C= \Omega_{A} \otimes \Omega_{B} \otimes \Omega_{C}$ decomposes in terms of local CPTP maps $\Omega_i$. If a state is not of the above form, it has tripartite GNME. In the case of 4 parties, we allow for tripartite resources such as $\sigma_{ABC}$, and so forth. 
We note that while the resource states $\sigma_{\#}$ can be of arbitrary size, the map $C$ gives a state in the original Hilbert space of $\rho$. This potential for infinite resources is a double-edged sword: it makes the set of network states both powerful and hard to characterize. Before developing tools to certify/quantify GNME, let us first discuss how it directly targets beyond-area-law multiparty entanglement.  

As we explained above, the GME encoded in sufficiently large subregions of a ground/Gibbs state of local Hamiltonians arises at leading order from interface bipartite entanglement. So given the density matrix of the $k$ subregions $\rho$, owing to the infinite resource potential, we can find a network state $\rho_{\rm net}$ that correctly reproduces the local interface entanglement, and thus the GME area law of $\rho$. Given the network approximant, a simple idea would then be to evaluate the difference $\mE(\rho) - \mE(\rho_{\rm net})$ to quantify GNME. Below we present concrete methods to get network approximants, and other approaches to quantify GNME in arbitrary states. 
\paragraph{Quantifying GNME.}
{\bf Inflation}---Quantifying GNME remains a major challenge, especially for generic mixed states. Partial results exist in restricted settings: network states without shared randomness~\cite{aaberg2020semidefinite,kraft2021quantum,Kraft2021coherence}, and highly symmetric families~\cite{Hansenne2022,luo2021new}. For general states, several SDP tests have been proposed, including the covariance matrix method with purity constraints~\cite{Xu2023Cov}, which applies mainly to nearly pure states. The most stringent test to date is the inflation technique~\cite{navascues2020genuine}, which certifies GNME through SDP, but suffers from poor scalability with system size. 
Intuitively, if a state $\rho$ can be prepared by a quantum network, hypothetically there exists an ``inflated'' state $\gamma$ that can be prepared by a larger network built from multiple copies of the original network. We can impose consistency constraints on the inflated state $\gamma$, and failure to satisfy these constraints falsifies the network hypothesis, thereby certifying GNME.
Building on the framework of Ref.~\cite{navascues2020genuine}, we leverage symmetries to efficiently implement higher-order inflation algorothms (up to 3rd order) and analyze 4-party GNME (see SM Sec.\ref{sup:certifyGNME}), thereby improving the certification windows.

{\bf Geometric distance}---We introduce a flexible convex-optimization framework to estimate GNME by evaluating the geometric distance between a given state and the set of network states.
The set of network states is convex, and so we can exploit methods to find the closest point in the network set to a given target state $\rho$. 
In practice, we work with a smaller set strictly contained within the network state: the unitary quantum network (UQN), which is formed by a convex hull of pure network states. 
These pure network states can be prepared by two-layer unitary circuits, where the first layer distributes the resources through $(k-1)$-partite unitaries and the second layer performs local unitary operations. For example, biseparable states are generated by depth-1 circuits shown in Fig.~\ref{fig:network}d, and a large class of network states are generated by the depth-2 brickwork circuits shown in Fig.~\ref{fig:network}e, which we call $B_2$. We then compute the geometric entanglement, which is the geometric distance of a given state to UQN: 
\begin{equation}
    D(\rho) = \min_{\sigma \in \rm UQN} d(\rho, \sigma) \end{equation}
where we use the Hilbert-Schmidt distance defined by the Frobenius norm $d(\sigma, \rho)=\sqrt{\text{tr}(\rho-\sigma)^2}$. Note that $D\geq D_{\rm net}$ upper bounds the distance to the full network set (see SM Sec.~\ref{sup:properties_GNME_dist} for the properties of $D_{net}$). We estimate $D$ using a variant of the Gilbert algorithm~\cite{brierley2016Gilbert,Pandya2020,Gilbert1966}, which iteratively searches for the point within a convex set that minimizes the distance to a given state, thus approximating the closest UQN (network) state to $\rho$. 
Moreover, the geometric distance can be used to rigorously certify that a noisy mixture of $\rho$ lies within the network set, according to the
Gilbert criterion~\cite{Shang2018} (see SM Sec.\ref{sup:certifyNet}).

\paragraph{GNME in standard states.}


\begin{table}[t]
    \centering
    \begin{tabular}{l|cc|c|cc}
    \hline
    \multirow{2}{*}{State} & \multicolumn{2}{c|}{Lower Bound} & \multirow{2}{*}{Estimate} & \multicolumn{2}{c}{Upper Bound} \\
    \cline{2-3} \cline{5-6}
     & Previous & This work & & Network & GMN \\
    \hline
    $\mathrm{GHZ}_3$ & 0.4366~\cite{neumann2025} & 0.3878 &  0.57 & \multicolumn{2}{c}{0.5715}   \\
    $W_3$            & 0.274~\cite{navascues2020genuine} & 0.3486 &  0.481 & 0.4824 & 0.5211 \\
    \hline
    $\mathrm{GHZ}_6$ & $0.4366^\dagger$ & 0.5121 &  0.711 & 0.7184 &  0.9140 \\ 
    $W_6$            & $0.274^\dagger$ & 0.4951 &  0.70 & 0.7280 & 0.8969 \\ 
    \hline\hline
    $D_{4,8}$        & - & 0.2408 & 0.72 &  0.7428  & 0.93 \\
    \hline
    \end{tabular}
    \tabnotetext{\dagger}{Inferred from their 3-qubit counterparts.}
    \caption{\textbf{GNME robustness thresholds.} 
    The threshold $p_c$ denotes the minimum noise for which the state becomes a network state.
    The lower bounds are obtained from the inflation technique, which certifies GNME, see SM Sec.\ref{sup:compare_lower_bound} for details on the previous bounds.
    The \emph{Estimate} is inferred from a linear extrapolation of the geometric distance. 
    The \emph{Network} upper bound marks the point above which the Gilbert criterion certifies network states, 
    while the GMN upper bound corresponds to the vanishing of genuine multiparty negativity.}
    \label{tab:bounds_gnme}
\end{table}

\begin{figure}[hbt!]
    \centering
    \includegraphics[width=\linewidth]{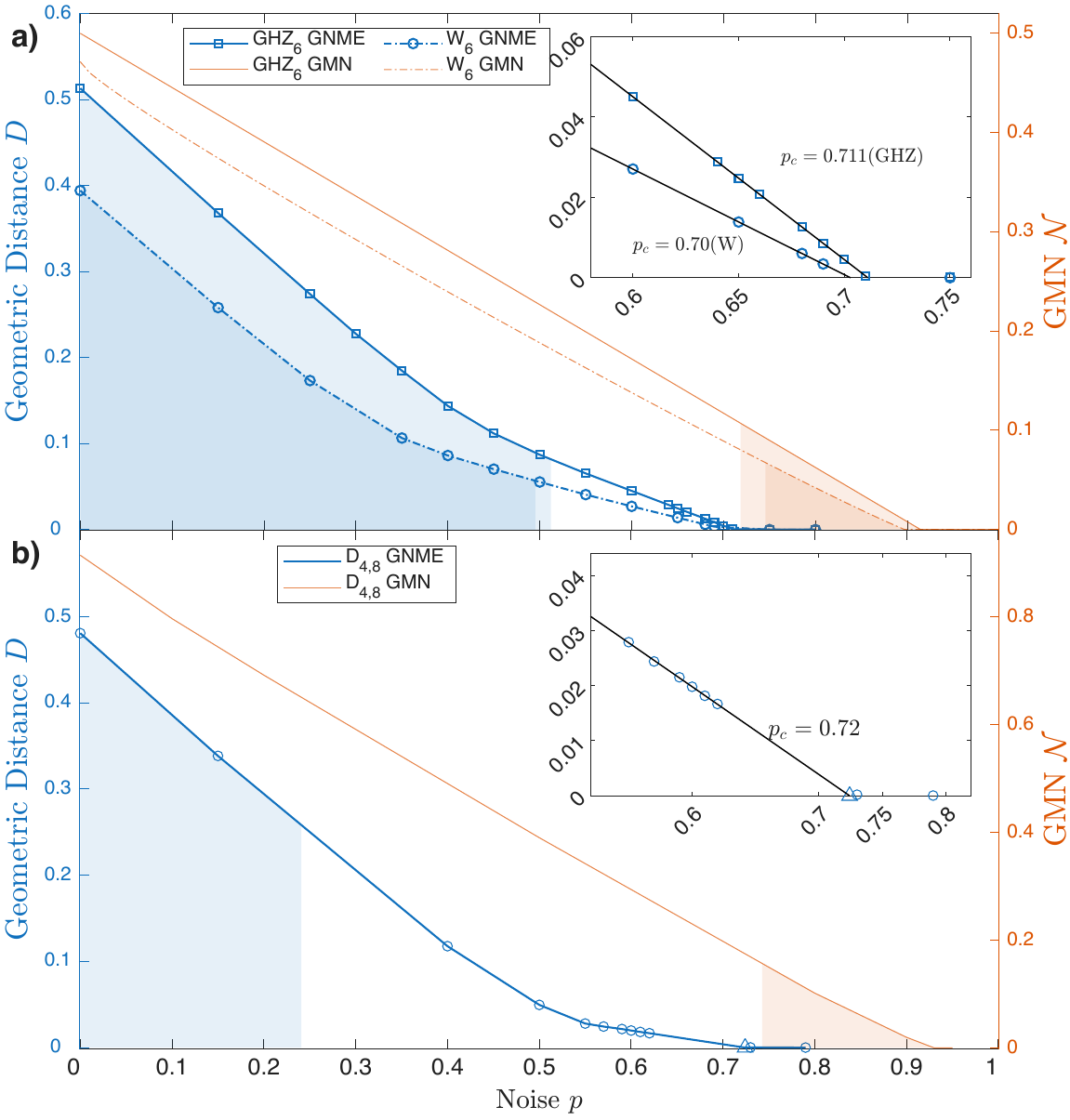}
    \caption{\textbf{GNME and GME vs white noise for pure states.} A pure state under white noise is given by $\rho=(1-p)\ket{\psi}\bra{\psi}+p\,\mathbb{I}/D$. 
    \textbf{(a)} Tripartite entanglement of $\mathrm{GHZ}_6$ (solid) and $W_6$ (dashed). 
    \textbf{(b)} Four-partite entanglement of the Dicke state $\ket{D_{4,8}}$. 
    Left axis: geometric distance to the unitary quantum network, right axis: GMN. The inset shows a linear fit used to estimate the threshold $p_c$ at which the distance vanishes. 
    The blue shaded regions mark the interval where GNME is certified. 
    The red shaded regions mark ranges of $p$ for which the state is GME but certified non-GNME.
    }
    \label{fig:ghz_dist}
\end{figure}

In Fig.~\ref{fig:ghz_dist}, we illustrate our methods on canonical pure GHZ, W and Dicke states of $n=6,8$ qubits mixed with white noise $(1-p)|\psi\rangle\langle\psi|+p\mathbb I/2^n$ where both $D$ and GMN $\mathcal N$ are plotted versus $p$; the case $n=3$ is shown in the Appendix. We find a substantially faster decline of $D$ compared to the GMN, leading to extended windows of finite GME but vanishing GNME (except for $\mathrm{GHZ}_3$). In particular, in all cases we observe a linear decrease of $D$ that precedes its vanishing. This linearity is geometrically well-motivated: the state describes a straight line and so we expect that its distance to a convex set (here, the set of network states) becomes linear in the affine parameter close to the entrance point.
This allows us to extrapolate to the point where $D=0$, providing an upper bound on the robustness: the value of $p$ at which a state becomes network. This is shown in the ``estimate'' column of Table~\ref{tab:bounds_gnme}. 
The identification of the entrance scaling provides results that would be otherwise impossible: brute force evaluation of $D$ becomes extremely challenging as one approaches the boundary of the convex set. 
Table~\ref{tab:bounds_gnme} further shows rigorous lower bounds on the GNME robustness obtained via inflation. Finally, the table provides rigorous upper bounds on the robustness by exploiting the separable ball around the identity, a method we call the Gilbert criterion~\cite{Shang2018}. A simple yet non-trivial result of this analysis is that the $W_3$ state mixed with $p=0.5$ white noise has finite GME but no GNME.

\begin{figure}[hbt!]
    \centering
    \begin{overpic}[width=\linewidth]{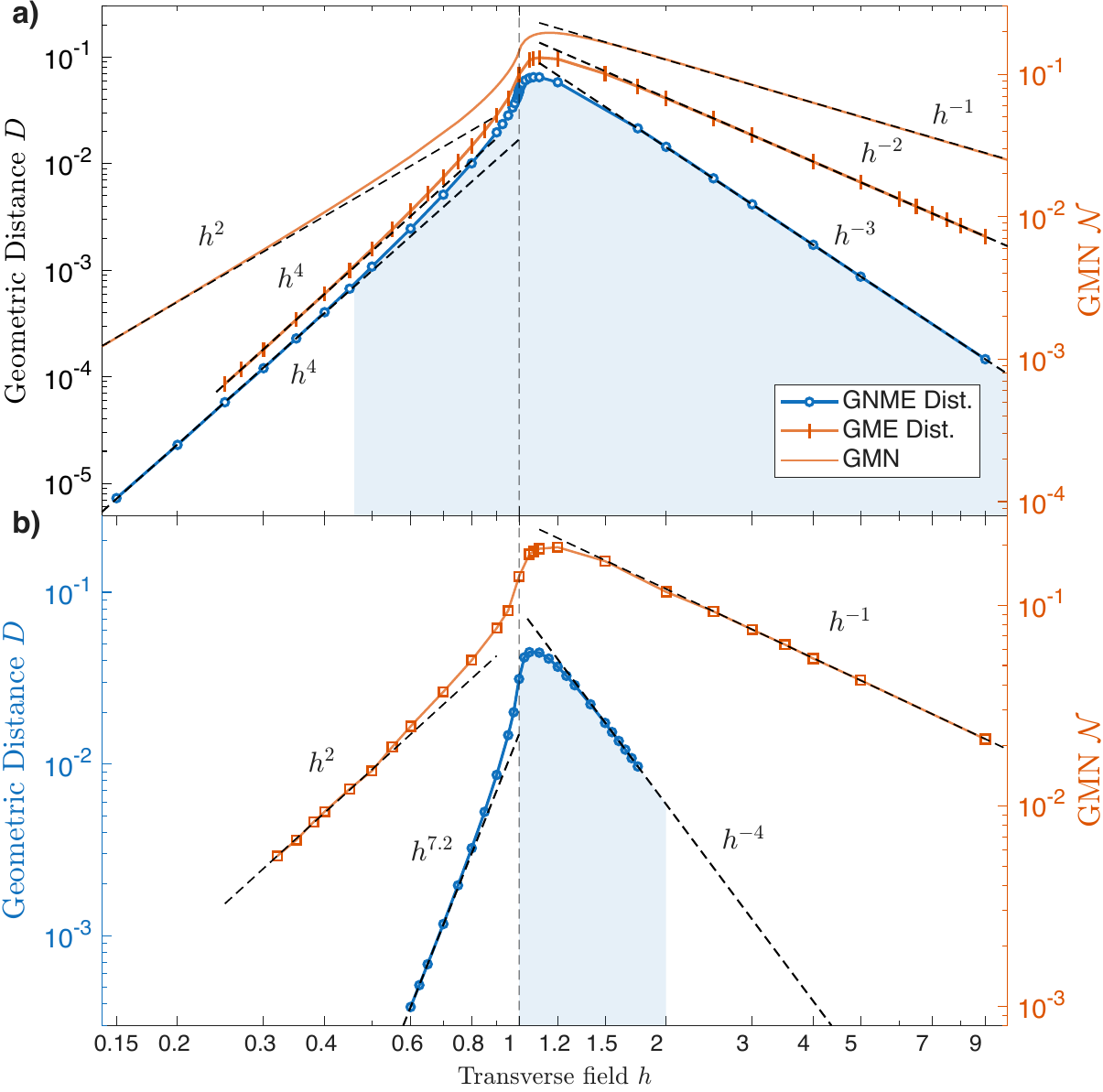}
        \put(10,89){\includegraphics[width=0.25\linewidth]{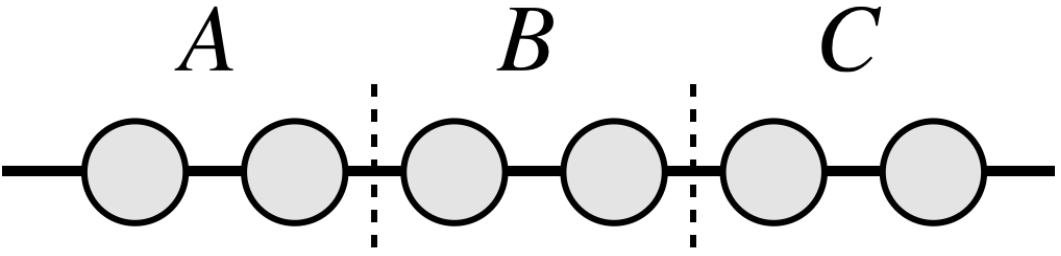}}
        \put(10,42){\includegraphics[width=0.28\linewidth]{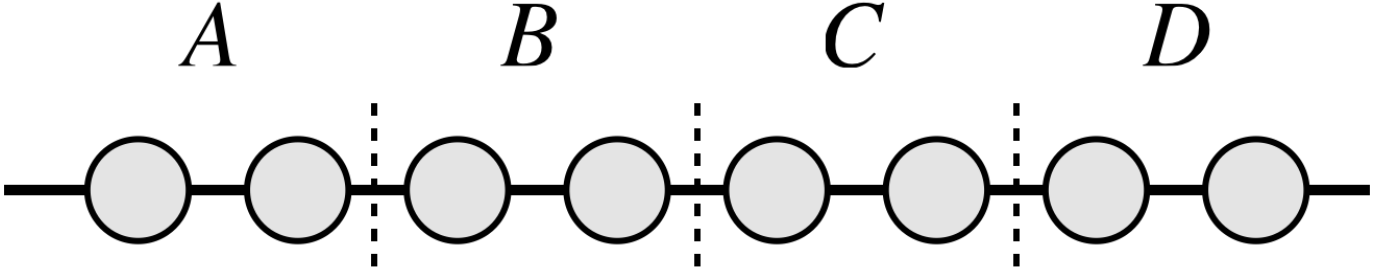}}
    \end{overpic}
\caption{\textbf{GNME and GME for adjacent spins in the 1d Quantum Ising model.}
  \textbf{(a)} Tripartite entanglement for 3 consecutive pairs: GNME distance (blue), GME distance (red, line marker), and GMN (red). GME distance provides an upper bound to GNME distance.  
  \textbf{(b)} Four-partite entanglement for 4 consecutive pairs.
  Left axis: geometric distance, right axis: GMN.
  The blue shaded regions, $[0.46,10]$ for 6 spins and $[1,2]$ for 8 spins, mark the interval where GNME is certified.}
    \label{fig:Ising_dist}
\end{figure}

\paragraph{GNME in Quantum Matter.}
We now turn to the study of GNME in states that arise in key quantum many-body systems. The first example is the transverse field Ising model:
\begin{equation}
    H =-\sum_{<i,j>} \sigma_i^x \sigma_j^x - h \sum_i \sigma^z_i
\end{equation}
where we have set the ferromagnetic (FM) coupling between nearest neighbors to unity. In the thermodynamic limit, this model has a quantum critical phase transition at $h_c$ from a FM to a paramagnet (PM).
 In 1d, the transition occurs at $h_c=1$, and the model is exactly soluble for all $h$ and temperatures via fermionization~\cite{Pfeuty1970, Fagotti2013RDM}; we exploit this to obtain the exact RDMs for subregions of up to 8 spins. Fig.~\ref{fig:Ising_dist} shows the $h$-dependence of the tripartite GNME and GME quantified via $D$ for three neighbouring pairs, as well as the GMN. We see that $D$ has a much more pronounced peak at the transition. We quantify this by identifying the small- and large-$h$ scaling of $D$, where the power for $D$ ($\mathcal{N}$) is 4 (2) and $-3$ ($-1$), respectively. 
 The GME distance stays clearly above $D$, with the same exponent in the FM but slower decay in the PM.
 The GMN exponents match those of bipartite entanglement between $A$ and $BC$ (SM Fig.~\ref{fig:Ising_neg}), in accordance with our discussion regarding the area law.
 Furthermore, we identify a logarithmic divergence of the derivative $dD/dh = -0.205\log|h-h_c| + \cdots$, which matches the critical scaling predicted for general entanglement measures in the Ising chain~\cite{Wang2025entanglement}. This shows the quality of our optimization for $D$. A log-divergence also occurs for $\mathcal{N}$ (SM Fig.~\ref{fig:Supplement_Ising_h_scaling}), in agreement with previous results for three consecutive spins~\cite{Hofmann2014Scaling}. 
Going beyond $D$, we use inflation to certify the presence of GNME for a large window, denoted in blue in Fig.~\ref{fig:Ising_dist}a. 
We then turn to the more challenging task of evaluating the 4-party $D$ of four adjacent pairs, as shown in Fig.~\ref{fig:Ising_dist}b. We observe an even sharper peak near the transition, and the same exponent at large $h$. The scaling in the FM is more challenging to converge. Inflation certifies the presence of GNME in a window (blue) that includes $h_c$. Based on a fate of entanglement argument~\cite{parez2026fate}, we expect GNME to be present for all $h$ since the asymptotes $h=0,\infty$ are at the boundary of the network set (see SM section~\ref{supp:boundary_proof}). Future improvements to inflation could allow broader certification. 

\begin{figure}[hbt!]
    \centering
    \begin{overpic}[width=\linewidth]{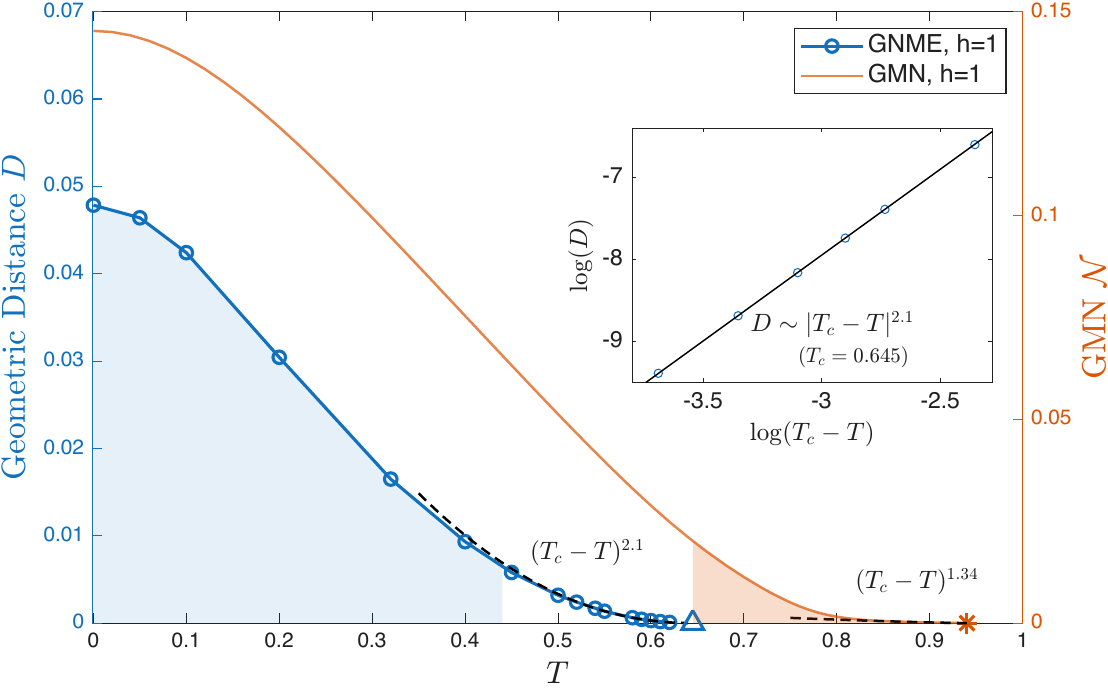}
        \put(30,51){\includegraphics[width=0.3\linewidth]{Figures/Ising_222_subregion.pdf}}
    \end{overpic}
    \caption{\textbf{Thermal GNME and GME in the quantum Ising model.} Left axis: geometric distance $D$, right axis: GMN $\mathcal{N}$. 
    The inset shows a power-law fit for $D$ as $T$ approaches the death-of-GNME temperature $T_c$. 
    The blue shaded region $T\leq0.44$ marks the window where GNME is certified. 
    The red shaded region $[0.65,0.9407]$ indicates finite GMN but no expected GNME.
    See SM Fig.~\ref{fig:GMN_loglog} for the GMN fit.
    } 
    \label{fig:Ising_finiteT}
\end{figure}

\begin{table}[t]
    \centering
    \renewcommand{\arraystretch}{1.15}
    \begin{tabular}{l|c|c|c}
    \hline
    State & Subregion & $\mathcal{D}$ & $\mathcal{N}$\\
    \hline
    GHZ$_6$ & & $0.5130~\certGNME$ & $0.5$\\ 
    GHZ$_{3}\otimes$GHZ$_3$ & & $0.5103~\certGNME$ & $1.5$\\
    W$_6$ & & $0.3920~\certGNME$ & $0.4714$\\ 
    \hline
    \multirow{2}{*}{Ising Chain QCP} & 12|34|56 & $4.78\times10^{-2}~\certGNME$ & 0.1453\\
     & $12\cdot34\cdot56$ & $9.0\times10^{-5}$ & 0.0021\\
    \hline
    Ising Square QCP & \multirow{2}{*}{
    \scalebox{0.9}{$
        \begin{array}{c|c|c} 
            1 & 3 & 5 \\[-3pt] 
            2 & 4 & 6
        \end{array}
    $}} & $1.2\times10^{-2}$ & 0.1093\\
    RVB Square & & $3.1\times10^{-3}$ & 0.0545\\
    \hline
    Kitaev Honeycomb & \multirow{2}{*}[1.8pt]{\hexagon[0.3]{\tiny}} & $0^{\dagger}$ & 0.2044\\
    Kagome CSL &  & $0^{\dagger}$ & 0.0252\\
    \hline
 \end{tabular}
 \tabnotetext{\checkmark}{certified GNME}
 \tabnotetext{\dagger}{zero within the precision of the optimization.}
 \caption{\textbf{GNME and GME for representative subregions from 1d and 2d many-body systems.} 
 Each subregion consists of three pairs of spins labelled $12$, $34$, and $56$. 
The table compares the geometric distance to the unitary network set $\mathcal{D}$ and the genuine multiparty negativity $\mathcal{N}$.
 }\label{tab:dist}
\end{table}


We next study three pairs at $h_c$ but with a separation of 1 site, see Table~\ref{tab:dist}. Separation makes $D$ decay much faster compared to $\mathcal{N}$, showing limited GNME for non-adjacent parties. We next analyze the fate of GNME at finite temperature by using the three adjacent pairs at $h_c$ as our testbed. In Fig.~\ref{fig:Ising_finiteT} we observe the rapid decline of $D$ compared to $\mathcal{N}$. By fitting the scaling of $D$ near the entrance point of the UQN, we predict a broad temperature window with GME but no GNME : $0.645<T<0.9407$. This fits with the overarching conclusion that the shorter-range GME is more prevalent in the phase diagram, with the immediate vicinity of the critical point showing the strongest GNME.

We then study the 2d Ising model at $h=3$ (near its critical coupling) on the 2d a $6\times5$ square lattice via exact diagonalization (ED). For 3 adjacent pairs of spins forming a $3\times2$ rectangle, we find a finite but reduced $D$ in agreement with heuristic expectations from monogamy: the higher connectivity distributes entanglement (including the network-irreducible type) more collectively, a property previously found for GME~\cite{Lyu2025multiparty}. Another quantum critical state is the nearest-neighbour resonating valence bond (RVB) state of singlets on the $6 \times 6$ square lattice studied via an exact wavefunction with the same subregion geometry. We find a finite but reduced value of $D$ compared to the above case.
Finally, we examine two quantum spin liquid ground states. The first one is the paradigmatic Kitaev model on the honeycomb lattice~\cite{Kitaev2006} that hosts gapless Majorana fermions, which we study exactly~\cite{lyu2025loop}. We search for tripartite GNME between three adjacent pairs on a hexagon. Although the GME is large, surprisingly, we find $D=0$ within the precision of our optimization. We can thus say with high probability that the state is a network state. 
We arrive at the same conclusion for the chiral spin liquid on the Kagome lattice (36-site ED) stabilized by a spin-chirality interaction~\cite{Bauer2014}. These findings shed new light on the entanglement structure of spin liquids, which was recently found to contain GME only in subregions forming closed loops~\cite{lyu2025loop}: Microscopic loopy subregions do not seem to carry truly collective entanglement. 

\paragraph{Conclusion.}
We have seen that genuine network multiparty entanglement provides a systematic way to capture truly collective entanglement. 
By contrast, in the limit of large subregions of ground/Gibbs states, the GME obeys an area law due to local bipartite entanglement across each interface. GNME systematically removes such a contribution, and more generally informs about the degree to which a state cannot be prepared by a network using bipartite resources (in the tripartite case). We have sharpened inflation methods to certify GNME, and introduced a Gilbert approach that can provide strong bounds on GNME. The methods were benchmarked on canonical states mixed with white noise, and used to analyze microscopic subregions of many-body systems. In all cases, GNME is substantially suppressed compared to GME, and even vanishes in paradigmatic spin liquids. 

Going forward, our ideas could be applied to better understand GNME and the related circuit-complexity of many-body systems, including topological phases and a broader range of quantum spin liquids, specifically to identify when GNME becomes finite. 
Furthermore, it would be interesting to apply these methods to systems out of equilibrium, such as in quantum quenches and dynamical phase transitions.
Finally, since our methods suffer from poor scalability with the dimension of the system, a next step will be to extend our analysis to larger subregions.


\section*{Acknowledgements}
We thank Elie Wolfe and Robert Spekkens for fruitful discussions. W.W.-K.\/ and L.L.\/ are supported by a grant from the Fondation Courtois, a Chair of the Institut Courtois, a Discovery Grant from NSERC, and a Canada Research Chair. Research at Perimeter Institute is supported by the Government of Canada through the Department of Innovation, Science and Economic Development Canada and by the Province of Ontario through the Ministry of Research, Innovation and Science.

\bibliographystyle{longapsrev4-2}
\bibliography{bibtex}

\clearpage
\onecolumngrid

\centerline{ \textbf{\large  Supplemental Materials for Network-Irreducible Multiparty Entanglement in Quantum Matter} }

\subsection{Algorithms for Genuine Network Multiparty entanglement}\label{sup:AlgorithmGNME} 
This section introduces algorithms used to quantify genuine network multiparty entanglement. We will discuss the Gilbert algorithm and how it can be used to certify that a state is a unitary network state. We will also discuss the inflation technique, which can certify when a state is GNME. \\

\begin{figure}[hbt!]
    \centering
    \includegraphics[width=0.5\linewidth]{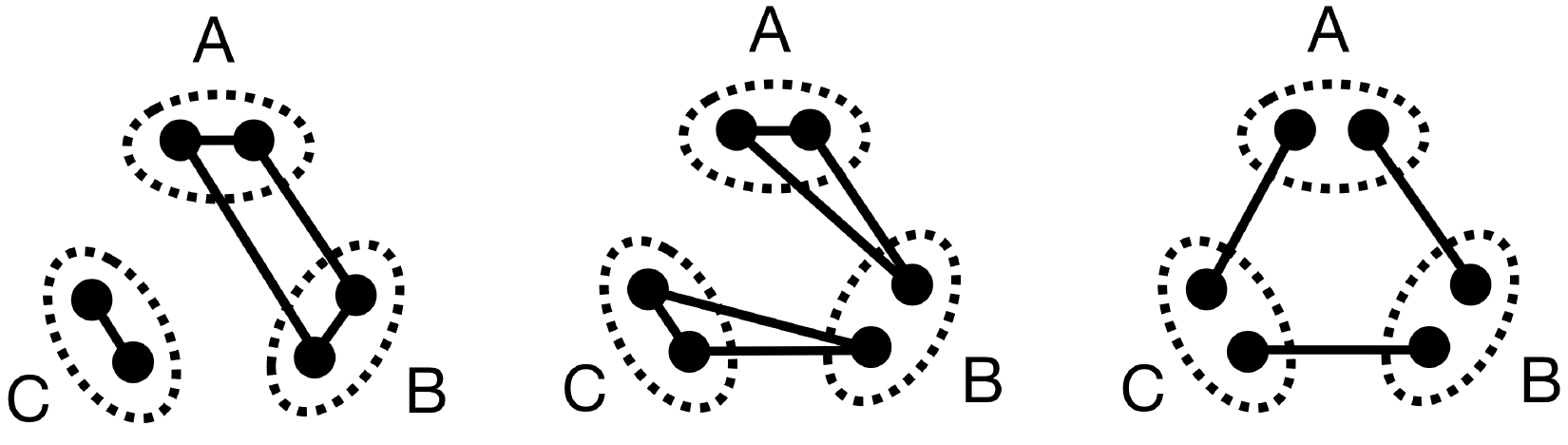}
    \caption{\textbf{Tripartite unitary networks}.}
    \label{fig:net3_diagrams}
\end{figure}

\begin{figure}[hbt!]
    \centering
    \includegraphics[width=0.5\linewidth]{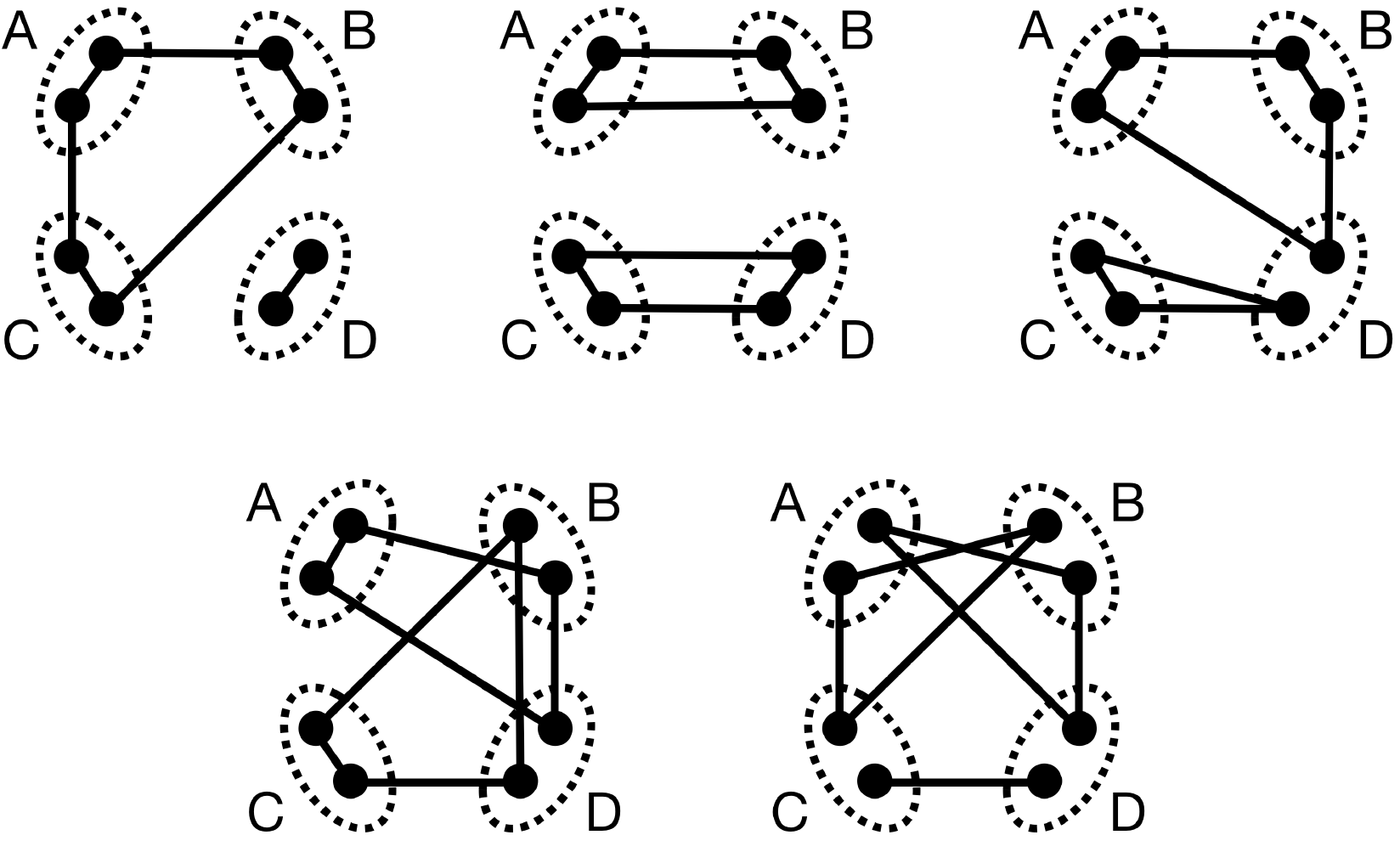}
    \caption{\textbf{Four-partite unitary networks}.}
    \label{fig:net4_diagrams}
\end{figure}

We begin by defining the set of unitary quantum network (UQN), which is an inner approximation of the set of network states. 
The extremal points of the UQN are pure network states which can be prepared by a two-layer quantum circuit: the first layer prepares the $(n-1)$-partite resource states, and the second layer performs local unitary gates within every party. The convex hull of these pure states form the set of \textbf{unitary quantum network}, which takes the form 
\begin{equation}
    \rho_{\text{u-net}}=\sum_\lambda p(\lambda) \ket{\psi_\lambda}\bra{\psi_\lambda}
\end{equation}
where $\ket{\psi_\lambda}$ is a pure unitary network state. For three parties, the two-layer brickwork circuit is shown in Fig.~\ref{fig:network} (e), and the circuit state is $\ket{\psi_\lambda}=U_A(\lambda) \otimes U_B(\lambda) \otimes U_C(\lambda) \left(\ket{s_{AB}}\otimes \ket{s_{BC}}\otimes \ket{s_{CA}}\right)$ where $\ket{s_{XY}}$ are bipartite resource states. 
In addition, any biseparable pure state which can be prepared in the same circuit, such as $\ket{\psi_\lambda}=\ket{s_{AB}}\otimes \ket{s_{C}}$ as shown in Fig.~\ref{fig:network} (d), is also a pure unitary network state. 
The four-partite unitary network states can be similarly constricted:
$\ket{\psi_\lambda}=U_A(\lambda) \otimes U_B(\lambda) \otimes U_C(\lambda) \otimes U_D(\lambda) \left(\ket{s_{ABC}}\otimes \ket{s_{ABD}}\otimes \ket{s_{ACD}}\otimes\ket{s_{BCD}}\right)$.
A list of all distinct 3- and 4-party unitary network states is shown in Fig.~\ref{fig:net3_diagrams} and Fig.~\ref{fig:net4_diagrams}. Connected parts represent that they are prepared by the same resource state. For three parties, this comes down to three types of unitary networks, unique up to any permutations of the three parties. The first network represents biseparable states, while the rest two represent networks with bipartite resources. For four parties, we allow for both bipartite and tripartite resources, resulting in five distinct unitary networks.

\subsubsection{Geometric Distance}\label{sup:dist}
Here, we define the geometric distance to the set of network states and describe how it is estimated using the Gilbert algorithm.
The geometric entanglement is defined as~\cite{Vedral1997geoent} 
\begin{equation}
    \mathcal{D}_{\text{net}}=\min _{\sigma \in \text{net}} d(\rho, \sigma)
\end{equation}
where $d\left(\rho, \sigma\right)=\sqrt{\operatorname{Tr}[\left(\rho-\sigma\right)^2]}$ is the Hilbert-Schmidt distance between two quantum states.
A direct evaluation of this distance is difficult, as generic network states are not decomposable into pure network state components (see SM of ~\cite{Hansenne2022}). This is the consequence of the fact that network states can come from local channels that map the resource state to a lower-dimensional Hilbert space. 
To make progress, we instead consider the distance $\mathcal{D}$ to the set of unitary quantum networks, which upper bounds $\mathcal{D}_{\text{net}}$:
\begin{equation}
    \mathcal{D}=\min _{\sigma \in \text{u-net}} d(\rho, \sigma),
\end{equation}
We can estimate an upper bound of this distance using the Gilbert Algorithm~\cite{Gilbert1966,Pandya2020}, described as follows: Given an input density matrix $\rho$, it adaptively finds $\rho_1$ within the convex set which minimizes the squared Hilbert-Schmidt distance $d^2\left(\rho, \rho_1\right)=\operatorname{Tr}[\left(\rho-\rho_1\right)^2]$ by the following algorithm
\begin{enumerate}
    \item \textbf{Initialize}: Select a mixed state inside the convex set as the initial $\rho_1$. 
    \item \textbf{Heuristic}: Solve the following sub-optimization problem: find a pure state $\ket{\psi}$ on the boundary of the convex set, which maximizes the expectation $\bra{\psi}\rho_1-\rho\ket{\psi}$. 
    This step should be repeated until a positive expectation is found (preselection). We then set $\rho_2=\ket{\psi}\bra{\psi}$.
    \item \textbf{Line Search}: Solve the quadratic problem by finding the optimal $p$ which minimizes the distance $\operatorname{Tr}\left(\rho-p \rho_1-(1-p) \rho_2\right)^2$. If the solution $p\in[0,1]$, update $\rho_1\leftarrow p \rho_1+(1-p) \rho_2$.
    \item go to step 2 and repeat, until the distance $D(\rho,\rho_1)$ converges. 
\end{enumerate}
An additional improvement on the Gilbert algorithm can be made by recording the pure states generated in step 2 (Gilbert with Memory~\cite{brierley2016convex}). Let $A$ denote the a $D^2 \times (M+1)$ matrix whose columns are vectorized RDMs, at the $k^{th}$ iteration, 
\begin{equation}
    A = [\text{vec}(\rho_{1}^{(k)}), \;\text{vec}(\ket{\psi_{k-M+1}}\bra{\psi_{k-M+1}}), \;...\; ,\text{vec}(\ket{\psi_{k-1}}\bra{\psi_{k-1}}),\;\text{vec}(\ket{\psi_{k}}\bra{\psi_{k}})]
\end{equation}
where the first column is the density matrix $\rho_1$, the rest $M$ columns are pure states generated in step 2. Step 3 is replaced by a simplex search, where we look for a convex mixture of all RDMs in the memory matrix $A$ that minimizes the distance to the input state $\rho$. Specifically, we find $x$ which minimizes
\begin{equation}
    \min_x ||A x-\text{vec}(\rho)||,\quad x\geq0, \quad \sum_i x_i=1
\end{equation}
and then update $\rho_1=Ax_*$. This optimization is a quadratic programming problem with linear constraints, so it can be efficiently solved. 
In practice, the Gilbert algorithm gives a good upper bound on the geometric distance. However, the upper bound does not directly certify the existence or absence of GNME, additional criteria are needed.

\subsubsection{Certifying Network States}\label{sup:certifyNet}
In this section, we outline the constructive algorithms used to certify whether a given mixed state lies inside the network set. The criterion introduced in Ref.~\cite{Shang2018}, which we call the \textbf{Gilbert criterion}, converts the geometric distance in the Gilbert algorithm into a threshold for white noise robustness. 

\begin{figure}[hbt!]
    \centering
    \includegraphics[width=0.5\linewidth]{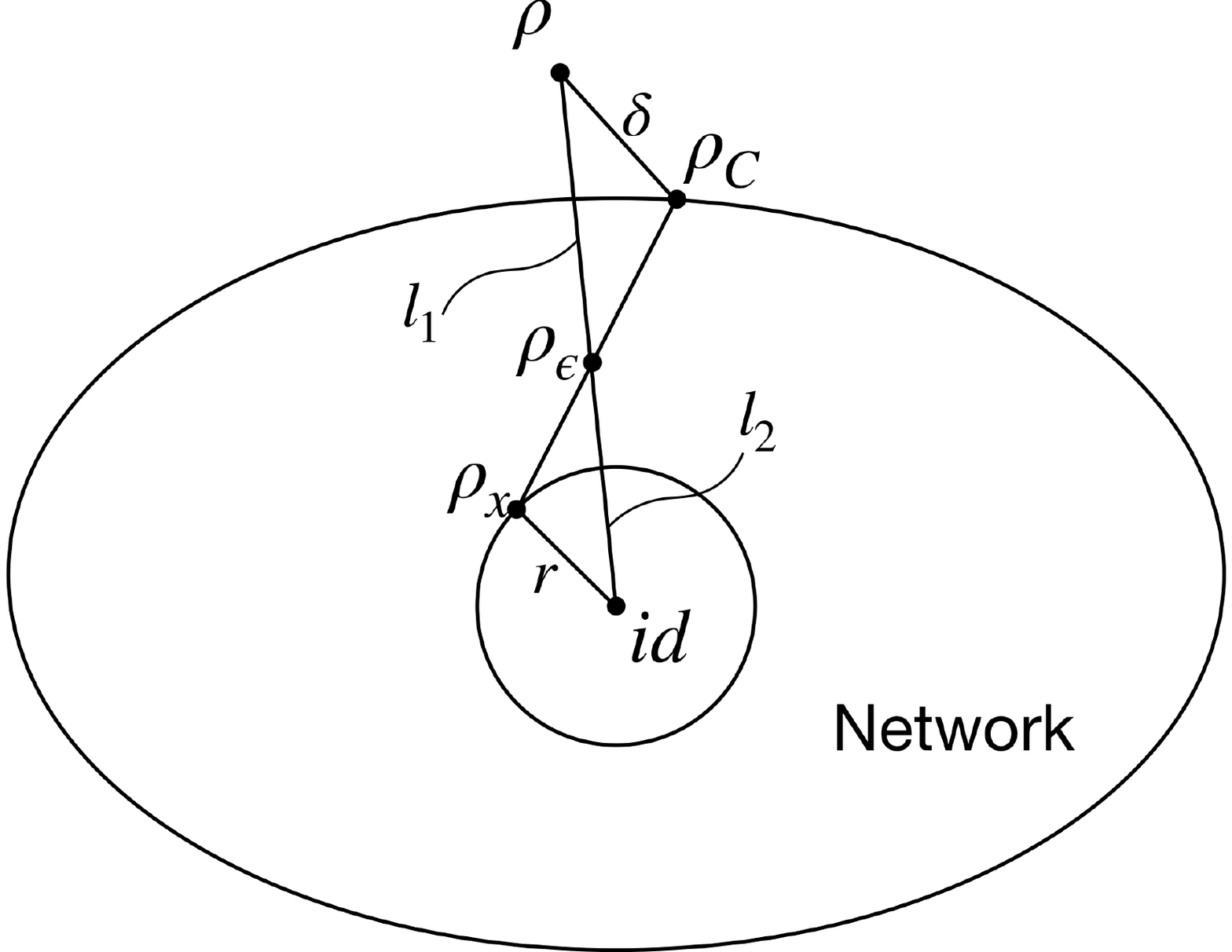}
    \caption{\textbf{Geometric discription of the Gilbert criterion}. Given a state $\rho$ and the closest network state $\rho_C$, we can certify $\rho_\epsilon=(1-\epsilon)\rho+\epsilon \;id$ to be a network state, where $\epsilon=\delta/(r+\delta)$. Here $r=1/\sqrt{D(D-1)}$ is the maximal radius around the maximally mixed state $id=I/D$ which is separable~\cite{Gurvits2003sepball} and $\delta=||\rho-\rho_C||$. Similarly, $l_1,l_2$ are Hilbert-Schmidt distance between $\rho_\epsilon$ and $\rho,\;id$. The noise $\epsilon$ is obtained by geometry $l_1/l_2=\delta/r$. $D$ is the Hilbert space dimension of $\rho$.}
    \label{fig:gilbert_cert}
\end{figure}

A geometric illustration of the algorithm is shown in Fig.~\ref{fig:gilbert_cert}. 
For the following geometric argument, it's enough to consider each density matrix as a vector in some finite-dimensional Hilbert space. 
Given state $\rho$, the Gilbert algorithm iteratively finds the (approximate) closest network state $\rho_C$ with distance $\delta=||\rho-\rho_C||$. On the other hand, there exists a ball around the maximally mixed states $id=I/D$ ($D$ is the dimension of $\rho$) in which the states are separable around any bipartition. This ball is inside the set of network states and has radius at least $r=1/\sqrt{D(D-1)}$~\cite{Gurvits2003sepball}. 
Define the state on the boundary of the ball $\rho_x= id + (r/\delta) \;(\rho-\rho_C)$ such that $\rho_x-id$ is parallel to $\rho-\rho_C$. 
The intersection between $\rho-id$ and $\rho_C-\rho_x$, $\rho_\epsilon$ defines a new state $\rho_\epsilon=l_2/(l_1+l_2) \rho + l_1/(l_1+l_2) \;id$, where $l_1=||\rho_\epsilon-\rho||$ and $l_2=||\rho_\epsilon-id||$. Since the two triangles are similar, we have $l_1/l_2=\delta/r$. We have, 
\begin{equation}
    \rho_\epsilon = \frac{r}{r+\delta} \rho + \frac{\delta}{r+\delta} id = \frac{r}{r+\delta} \rho_C + \frac{\delta}{r+\delta} \rho_x
\end{equation}
From the second equality, it's clear that $\rho_\epsilon$ is a network state, since it is a mixture of two network states. On the other hand, the first equality says that $\rho_\epsilon$ is simply $\rho$ mixed with noise $\epsilon=\delta/(r+\delta)$.
This gives the following criterion
\begin{criterion}
Given a $D$-dimensional quantum state $\rho$ with distance $d$ to the set of network states, if $d\leq\delta$, then $\rho_\epsilon=(1-\epsilon)\rho+\epsilon\;(I/D)$ is a network state, where $\epsilon=\delta/(\delta+1/\sqrt{D(D-1)})$.
\end{criterion}
In general, the above criterion gives a robustness threshold for any quantum state. However, if we wish to certify that a state is a network state, we should first "purify" the state by adding to the state $\epsilon(\rho-I/D)$ for some small $\epsilon$, and then run the Gilbert algorithm to get the closest network state to this new state. If the input state is non-full rank or has a vanishing smallest eigenvalue, as is the case for reduced states from 1d quantum Ising model, the purification cannot be done, and thus the certification does not apply. 

\subsubsection{Certifying Genuine network multiparty entanglement}\label{sup:certifyGNME}
\begin{figure}[hbt!]
    \centering
    \includegraphics[width=0.6\linewidth]{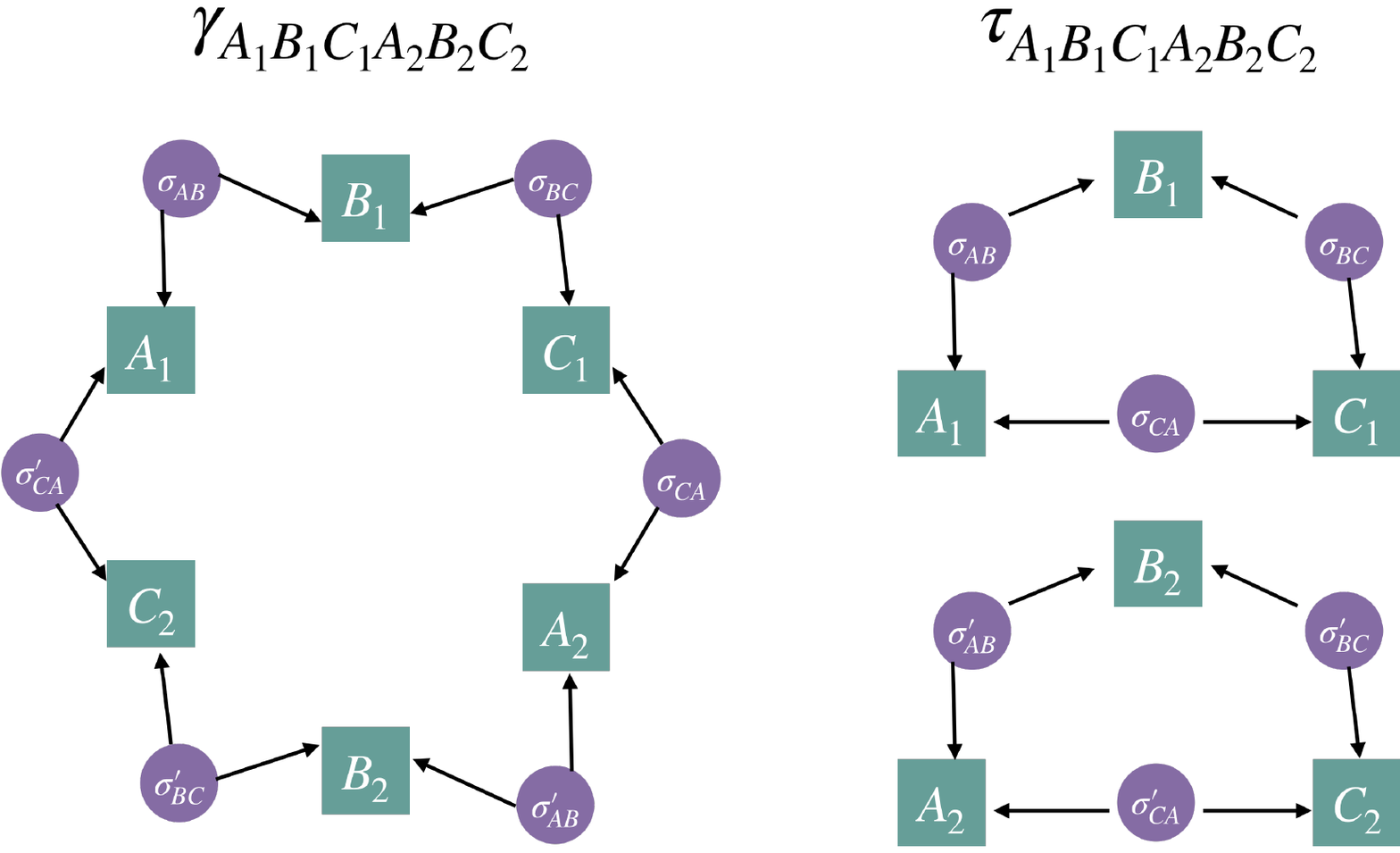}
    \caption{\textbf{Inflation level 2:} \emph{Ring inflation} on the left and two disjoint three-partite triangle networks (omitting the global shared randomness $\lambda$).
    We consider inflations with only two copies of each source.}
    \label{fig:inflation_triangle}
\end{figure}

\begin{figure}[hbt!]
    \centering
    \includegraphics[width=0.8\linewidth]{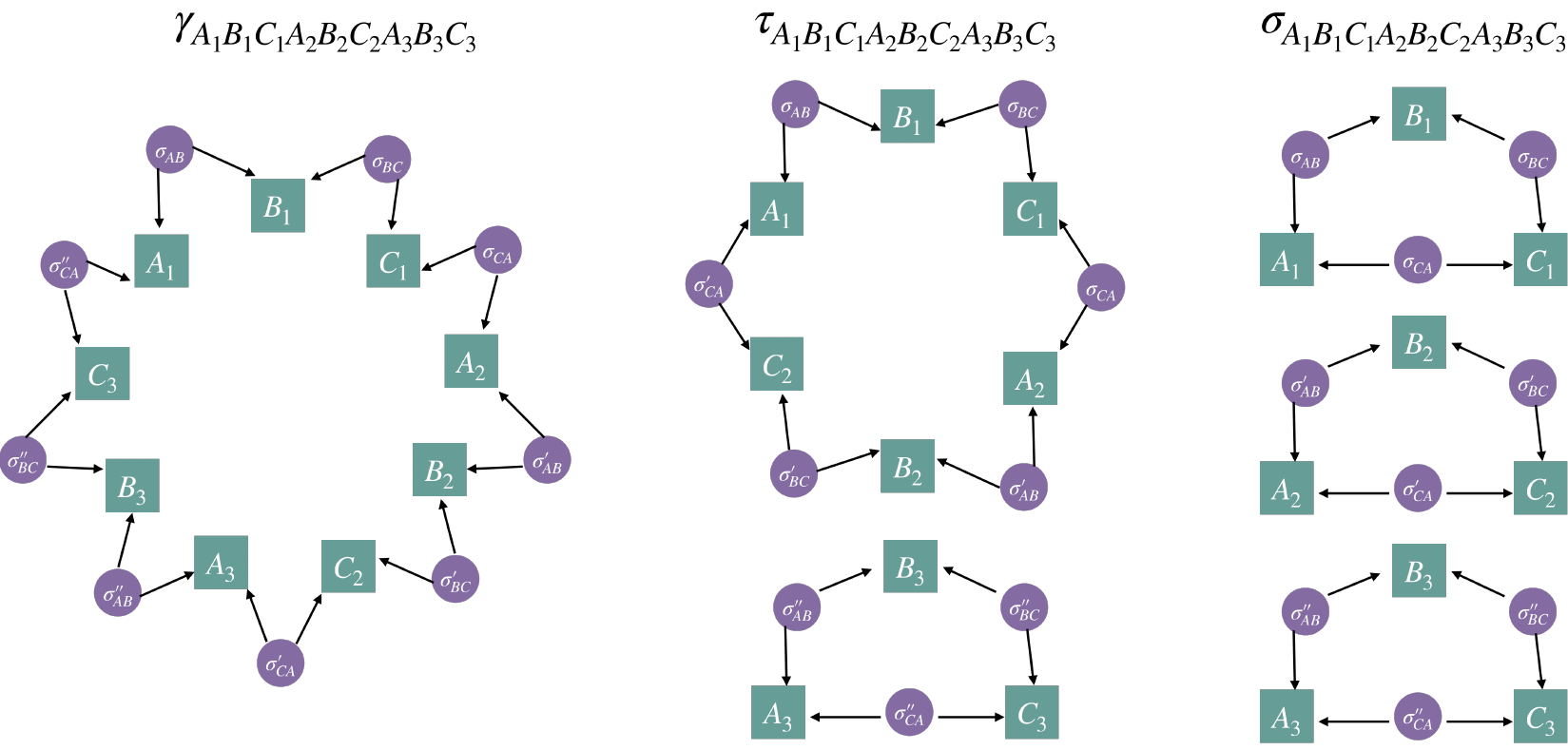}
    \caption{\textbf{Inflation level 3:} Inflations with three copies of each source. We have the \emph{9-ring inflation} on the left, the ring inflation and a disjoint triangle network, and three disjoint triangle networks on the right (omitting the global shared randomness $\lambda$).
    }
    \label{fig:inflation_level3}
\end{figure}

\begin{figure}[hbt!]
    \centering
    \includegraphics[width=0.6\linewidth]{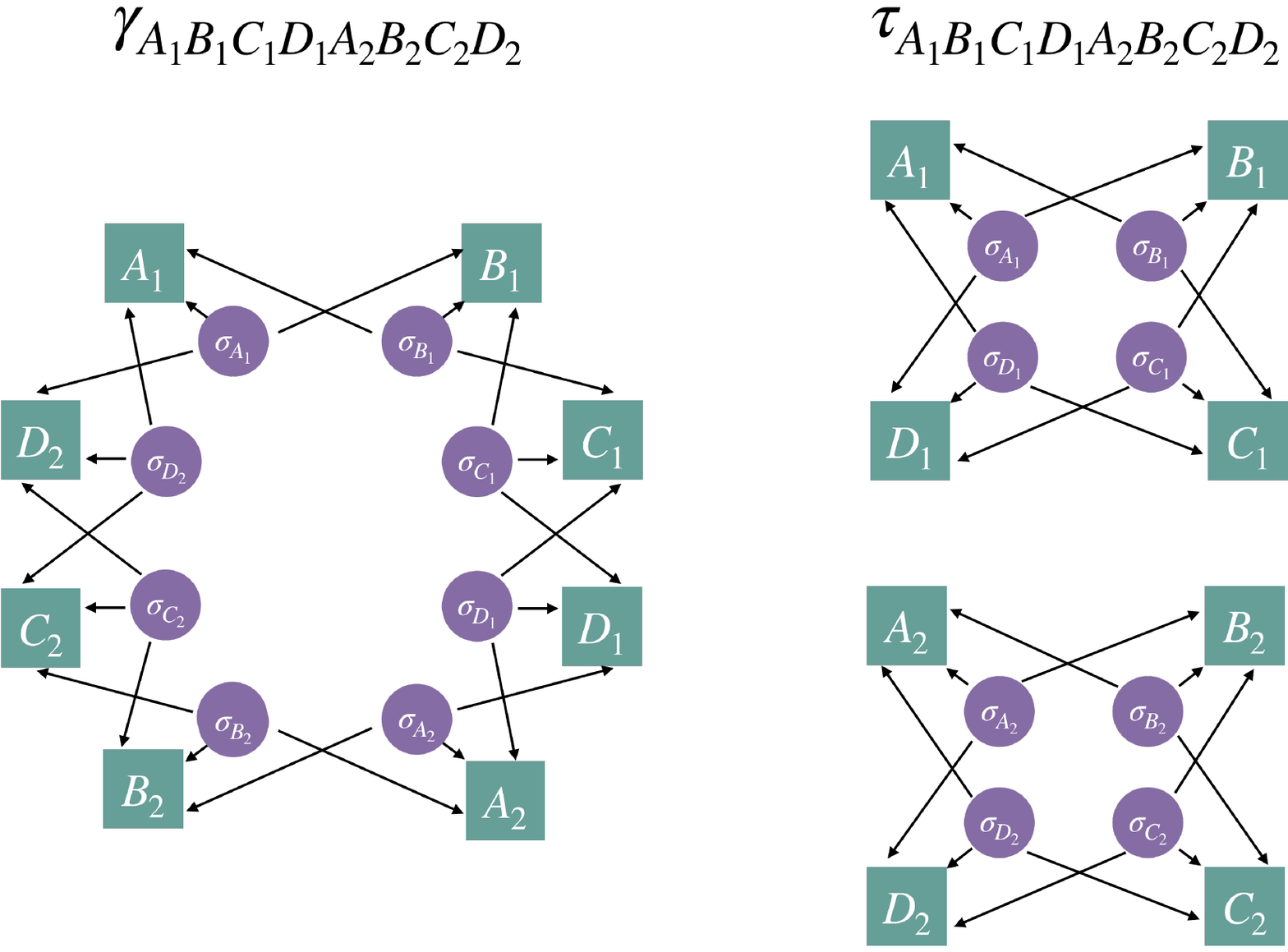}
    \caption{\textbf{Inflation level 2 for the four-partite scenario}: \emph{8-ring inflation} with tripartite sources of entanglement on the left and two disjoint \emph{tetrahedron networks} on the right, and we omit the global shared randomness $\lambda$.
    }
    \label{fig:inflation_tetrahedron}
\end{figure}

To certify GNME, we employ the inflation technique \cite{WolfeSpekkensFritz+2019,Wolfe2021inflation}, which provides an outer approximation to the network set given by a semidefinite program (SDP). To certify that a state lies outside the set of network states, a general strategy is to work with a tractable \emph{inflation}, i.e. a tractable outer approximation, and show that the state is not in this outer set. Directly characterizing the network set is difficult because network states may arise from distributing bipartite entanglement of \emph{unbounded} local dimension followed by arbitrary local operations. The framework of the inflation technique addresses this by imposing necessary constraints that any state generated by a given network structure must satisfy; e.g. given a GNME network structure, we consider a larger (or inflated) scenario that is composed of independent copies of the components of the original GNME network
$\sigma_{A_0B_0} $,  $\sigma_{B_1C_0}$, $\sigma_{C_1A_1}$, $\Omega^{\lambda}_{A}$, $\Omega^{\lambda}_{B}$ and $ \Omega^{\lambda}_{C}$, see Figs. \ref{fig:inflation_triangle},\ref{fig:inflation_level3},\ref{fig:inflation_tetrahedron} for some examples. The purple circles denote the original sources of entanglement, and the green squares represent CPTP maps of each node. After building these inflations, we impose some consistency constraints that any state generated by the original GNME network must satisfy, given by 

\begin{enumerate}
\item \textbf{Trivial conditions:} positivity and unit trace of each density operator considered.
\item \textbf{Matching marginals:} whenever a subset of nodes in the inflated network has the same structure as a set of nodes in the original GNME network, they have identical marginal states.
\item \textbf{Copy symmetries:} permutations that merely relabel identical copies leave the inflated state unchanged (e.g., swapping two identical blocks).
\item \textbf{Independence:} if two subsets of nodes in the inflated diagram are not connected, the joint state across that split must be unentangled (separable; often enforced via a positive partial transpose (PPT) check).
\end{enumerate}

and these constraints form a convex, SDP-representable outer approximation of the original GNME set. Violating these induced compatibilities certifies that the given state is not realizable by the original network. In prior work, inflation was used to upper bound the maximal network fidelity with target pure states (e.g., GHZ, W) via an SDP~\cite{navascues2020genuine}.



As a specific example, to certify that a \emph{mixed} state \(\rho\) is outside the set of triangle network states, we consider the ring inflation diagram shown in Fig.~\ref{fig:inflation_triangle}, which was considered in Ref.~\cite{navascues2020genuine}. We introduce two six-party inflated quantum states, $\tau_{A_1B_1C_1A_2B_2C_2}$ and $\gamma_{A_1B_1C_1A_2B_2C_2}$, where $A_i,B_i,C_i$ denote the $i$-th inflated copy of the original parties; The resource states $\sigma_{XY}$ and $\sigma_{XY}^\prime$ are identical copies of the same state. Following the constraints described above, we can construct the following robustness SDP:
\begin{equation}
\begin{aligned}
R(\rho_{ABC}) &= \max \; t \\
\text{subject to }\quad & t\geq0, \\
& \tau_{(A_1B_1C_1)} = t \rho_{ABC} + (1-t) I/D  \\
& \tau,\;\gamma\succeq0,\; \Tr(\tau) = \Tr(\gamma) = 1,  
 \tau_{21} =\tau, \;\gamma_{21} =\gamma, \\
& \gamma_{(A_1B_1A_2B_2)}=\tau_{(A_1B_1A_2B_2)},\; \gamma_{(B_1C_1B_2C_2)}=\tau_{(B_1C_1B_2C_2)}, \; \gamma_{(C_1A_2C_2A_1)}=\tau_{(C_1A_1C_2A_2)} \\
& \tau^{T_1}, \;\gamma_{(A_1B_1C_1B_2)}^{T_{B_2}}, \gamma_{(B_1C_1A_2C_2)}^{T_{C_2}},  \gamma_{(C_1A_2B_2A_1)}^{T_{A_1}} \succeq 0
\end{aligned}
\end{equation}
where $A\succeq0$ means square matrix A is semi-positive definite, reduced states are denoted by a subscript in parentheses listing the subsystems kept in the written order; for example,
\(
\tau_{(A_1B_1A_2B_2)} \;:=\; \Tr_{C_1C_2}\,\tau,
\)
which lives on $A_1\otimes B_1\otimes A_2\otimes B_2$ in that order. $\tau_{21}$ is a short-hand notation for the permuted state $\tau_{A_2B_2C_2A_1B_1C_1}$, and similarly for $\gamma$. $T_1$ denotes the partial transpose with respect to the subsystem $A_1B_1C_1$.
The separability criterion is relaxed to the PPT criterion on the indicated marginals, yielding an SDP-feasible outer approximation to the network set; these can be further tightened by replacing PPT with stronger tests such as PPT+symmetric extensions~\cite{Doherty2002ksymmext} on the relevant reduced states. The SDP program is modelled using the YALMIP toolbox~\cite{YALMIP2004} in Matlab, with the splitting conic solver (SCS)~\cite{Brendan2016scs}.
In the program above, the scalar \(t\) is the largest weight with which the target state \(\rho_{ABC}\) can appear in a white-noise mixture that admits a feasible inflation (i.e., satisfies all linear and PPT constraints). Hence,
\[
t_{\max} \;=\; \max\Big\{\, t\in[0,1] \;:\; t\,\rho_{ABC}+(1-t)\,\tfrac{I}{D}\ \text{has an inflation-consistent extension}\,\Big\}.
\]
If \(t_{\max}<1\), then \(\rho_{ABC}\) itself (\(t=1\)) does \emph{not} belong to the inflation set. Since every network state lies in the inflation set (\(\mathcal{N}\subseteq\mathcal{I}\)), this certifies that $\rho_{ABC}$ is not a network state; in other words, \(\rho_{ABC}\) possesses genuine network multiparty entanglement. 

A higher order inflation that we call level-3 inflation can be constructed by including three 9-party inflated states, as shown in Fig.~\ref{fig:inflation_level3}. The SDP is as follows:
\begin{equation}
\begin{aligned}
R(\rho_{ABC}) &= \max \; t \\
\text{subject to }\quad & t\geq0, \\
& \sigma_{(A_1B_1C_1)} = t \rho_{ABC} + (1-t) I/D  \\
& \gamma,\;\tau,\; \sigma\succeq0,\; \Tr(\gamma) = \Tr(\tau) = \Tr(\sigma) = 1,  \\ 
& \gamma_{231} =\gamma, \;
\tau_{213} = \tau, \; \sigma_{213}=\sigma_{132}=\sigma, \\
 &\tau^{T_3}, \; \sigma^{T_3} \succeq 0,\\
& \gamma_{(B_3 C_3 A_1 B_1 C_1 B_2 C_2)}=\tau_{(B_2 C_2 A_1 B_1 C_1 B_3 C_3)},\; 
\tau_{(B_1 C_1 B_2 C_2 A_3 B_3 C_3)}=\sigma_{(B_1 C_1 B_2 C_2 A_3 B_3 C_3)}, \\ 
& \gamma_{(A_1 B_1 C_1 A_2 B_2 A_3 B_3)}=\tau_{(A_1 B_1 C_1 A_2 B_2 A_3 B_3)},\;
\tau_{(A_1 B_1 A_2 B_2 A_3 B_3 C_3)}=\sigma_{(A_1 B_1 A_2 B_2 A_3 B_3 C_3)}, \\
&\gamma_{(C_3 A_1 B_1 C_1 A_2 C_2 A_3)}=\tau_{(C_2 A_1 B_1 C_1 A_2 C_3 A_3)},\;
 \tau_{(C_1 A_2 C_2 A_1 A_3 B_3 C_3)}=\sigma_{(C_1 A_1 C_2 A_2 A_3 B_3 C_3)},
\\
& \gamma_{(A_1 C_1 A_2 B_2 C_2 A_3 B_3)}^{T_{A_1}}, \gamma_{(B_1 A_2 B_2 C_2 A_3 B_3 C_3)}^{T_{B_1}}, \gamma_{(C_1 B_2 C_2 A_3 B_3 C_3 A_1)}^{T_{C_1}}\succeq 0, \\
& \gamma_{(A_1 B_1 C_1 B_2 C_2 A_3 B_3)}^{T_{A_1B_1C_1}}, \gamma_{(B_1 C_1 A_2 C_2 A_3 B_3 C_3)}^{T_{B_1 C_1 A_2}}, \gamma_{(C_1 A_2 B_2 A_3 B_3 C_3 A_1)}^{T_{C_1 A_2 B_2}}\succeq 0, \\
& \tau_{(A_1 C_1 A_2 B_2 A_3 B_3 C_3)}^{T_{A_1}}, 
\tau_{(B_1 A_2 B_2 C_2 A_3 B_3 C_3)}^{T_{B_1}}, 
\tau_{(C_1 B_2 C_2 A_1 A_3 B_3 C_3)}^{T_{C_1}} \succeq 0, \\
& \tau_{(A_1 C_1 A_2 B_2 A_3 B_3 C_3)}^{T_{C_1 A_2 B_2}}, 
\tau_{(B_1 A_2 B_2 C_2 A_3 B_3 C_3)}^{T_{A_2 B_2 C_2}}, 
\tau_{(C_1 B_2 C_2 A_1 A_3 B_3 C_3)}^{T_{B_2 C_2 A_1}} \succeq 0
\end{aligned}
\end{equation}
where states such as $\gamma_{231}$ denote the corresponding permuted state $\gamma_{A_2B_2C_2A_3B_3C_3A_1B_1C_1}$. We have matching marginals between both $\gamma,\tau$ and $\tau,\sigma$, and separability conditions apply to all three inflated states. 

For the 4-partite GNME, the resource states are three-partite states, so the network under consideration is a tetrahedron network, where each vertex is a party, and each face of the tetrahedron is associated with a tripartite resource. To inflate the tetrahedron network, we first notice that we can layout the four parties in a ring, as shown in Fig.~\ref{fig:inflation_tetrahedron} right panel, and then associate each source to its corresponding party. This way, each source is distributed to its own party and the neighbouring parties. 
We can then apply the same type of ring inflation as we did for the triangle network, resulting in the 8-party inflated diagram shown on the left panel. The compatibility constraints are similar to the triangle network, but it's important to note that
Marginal states such as $\gamma_{(A_1B_1C_1A_2B_2C_2)}$ are still connected by $\sigma_{D_{1/2}}$, so it does not agree with the corresponding marginal states of $\tau$. Only 4-party marginal states should agree, such as  $\gamma_{(A_1B_1A_2B_2)}=\tau_{(A_1B_1A_2B_2)}$. Similarly, the separability constraint for $\gamma$ does not apply to 6-party or 5-party marginal states, but applies to 4-qubit marginals $\gamma_{(A_1B_1C_1B_2)}^{T_{B_2}}$. The robustness SDP for 4-partite GNME is:
\begin{equation}
\begin{aligned}
R(\rho_{ABCD}) &= \max \; t \\
\text{subject to }\quad & t\geq0, \\
& \tau_{(A_1B_1C_1D_1)} = t \rho_{ABCD} + (1-t) I/D  \\
& \tau,\;\gamma\succeq0,\; \Tr(\tau) = \Tr(\gamma) = 1,  \tau_{21} =\tau, \;\gamma_{21} =\gamma, \\
& \gamma_{(A_1B_1A_2B_2)}=\tau_{(A_1B_1A_2B_2)},\; \gamma_{(B_1C_1B_2C_2)}=\tau_{(B_1C_1B_2C_2)}, \; \\
& \gamma_{(C_1D_1C_2D_2)}=\tau_{(C_1D_1C_2D_2)} \; 
\gamma_{(D_1A_2D_2A_1)}=\tau_{(D_1A_1D_2A_2)}
\\
& \tau^{T_1} ,\;\gamma_{(A_1B_1C_1B_2)}^{T_{B_2}}, \gamma_{(B_1C_1D_1C_2)}^{T_{C_2}},  \gamma_{(C_1D_1A_2D_2)}^{T_{D_2}},
\gamma_{(D_1A_2B_2A_1)}^{T_{A_1}}
\succeq 0
\end{aligned}
\end{equation}

For 4-party inflation of 8-qubit quantum states, the dimension for $\gamma$ and $\tau$ is $D=4^8=65{,}536$, which is memory intensive. However, we can use the property that GNME is monotonic under LOSR to lower bound the inflation robustness: If we apply a local channel (CPTP map) at each party which maps the state to a lower-dimensional quantum state, GNME cannot increase. For the Dicke state as well as the 8-spin Ising state, we first construct the reduced density matrix of each party ($d=4$), and then define a channel that projects each party onto an effective three-level subspace. Let $\{|{\psi_i}\rangle\}_{i=1}^{4}$ be the eigenbasis of the local density matrix, ordered by decreasing eigenvalue. We define the map
\begin{equation}
\begin{aligned}
K \;&=\; \sum_{i=1}^{3} \ket{i}\!\bra{\psi_i},\\
Q_i \;&=\; \frac{1}{\sqrt{3}}\,\ket{i}\!\bra{\psi_4},\qquad i=1,2,3,
\end{aligned}
\end{equation}
with channel
\begin{equation}
\Lambda(\rho)\;=\; K\rho K^\dagger + \sum_{i=1}^{3} Q_i \rho Q_i^\dagger.
\end{equation}
Trace preservation holds because
\begin{equation}
K^\dagger K + \sum_{i=1}^{3} Q_i^\dagger Q_i 
= \sum_{i=1}^{3} \ket{\psi_i}\!\bra{\psi_i} + \frac{1}{3}\sum_{i=1}^{3} \ket{\psi_4}\!\bra{\psi_4} = I_4.
\end{equation}
Moreover,
\begin{equation}
K K^\dagger + \sum_{i=1}^{3} Q_i Q_i^\dagger 
= \sum_{i=1}^{3} \ket{i}\!\bra{i} + \frac{1}{3}\sum_{i=1}^{3} \ket{i}\!\bra{i} 
= \frac{4}{3}\,I_3,
\end{equation}
so the maximally mixed state is preserved:
\begin{equation}
\Lambda\!\left(\frac{I_4}{4}\right) \;=\; \frac{1}{4}\,\Lambda(I_4) \;=\; \frac{1}{4}\cdot\frac{4}{3}\,I_3 \;=\; \frac{I_3}{3}.
\end{equation}
Consequently, any white-noise mixture maps to a white-noise mixture:
\begin{equation}
\Lambda\!\bigl((1-p)\rho + p\,I_4/4\bigr) 
= (1-p)\,\Lambda(\rho) + p\,I_3/3.
\end{equation}
This guarantees that GNME lower bounds obtained after the local dimension reduction (by LOSR monotonicity) are directly comparable to robustness thresholds defined with respect to white noise: if the original state $\rho$ has robustness threshold $p_c$, the reduced state $\Lambda(\rho)$ necessarily has $p_c' \le p_c$.

\paragraph{Numerical tip: symmetry-reduced parametrization.}
Inflation SDPs grow rapidly with dimension; a simple reduction is to build the \emph{swap symmetry} in from the start. 
Let \(D=\dim(ABC)\), so the inflated states \(\tau,\gamma\in\mathbb{C}^{D^2\times D^2}\) act on two identical blocks \((ABC)_1(ABC)_2\).
Let \(V\) be the swap on these blocks and define projectors
\[
P_{\pm}=\tfrac12\bigl(\mathbb{I}\pm V\bigr),\qquad 
P_\pm^2=P_\pm,\; P_+P_-=0.
\]
Choose isometries \(Q_\pm\) with orthonormal columns spanning \(\mathrm{range}(P_\pm)\) so that \(Q_\pm^\dagger Q_\pm=\mathbb{I}\) and \(P_\pm=Q_\pm Q_\pm^\dagger\).
Since \([\tau,V]=[\gamma,V]=0\), both states block-diagonalize in the \(\pm\) sectors:
\[
\tau \;=\; Q_+\,Y_+\,Q_+^\dagger \;+\; Q_-\,Y_-\,Q_-^\dagger,\qquad
\gamma \;=\; Q_+\,Z_+\,Q_+^\dagger \;+\; Q_-\,Z_-\,Q_-^\dagger,
\]
with \(Y_\pm,Z_\pm\succeq 0\) Hermitian decision variables. 
Trace constraints become
\[
\operatorname{tr}(\tau)=\operatorname{tr}(Y_+)+\operatorname{tr}(Y_-)=1,\quad
\operatorname{tr}(\gamma)=\operatorname{tr}(Z_+)+\operatorname{tr}(Z_-)=1.
\]
This parametrization \emph{automatically} enforces swap symmetry and reduces matrix sizes from \(D^2\) to
\[
n_+ \;=\; \tfrac{D(D+1)}{2},\qquad \text{and}\qquad
n_- \;=\; \tfrac{D(D-1)}{2}\qquad
\]
For the triangle network of six qubits (\(D=64\)), this yields \(n_+=2,080\), \(n_-=2,016\) (versus \(64^2=4,096\)) per block, significantly improving runtime and conditioning. 

\subsubsection{Proof that the $h=0$ and $h\to\infty$ limits lie on the boundary of network states}\label{supp:boundary_proof}

In the main text (see Fig.~\ref{fig:Ising_finiteT}), we utilized the inflation technique to certify the presence of Genuine Network Multiparty Entanglement (GNME) for a subregion of 6 consecutive spins in the 1d Transverse Field Ising Model (TFIM). Specifically, we achieved certification within the window $h \in [0.46, 10]$. Outside this window, while the inflation hierarchy employed could not definitively certify GNME due to convergence limitations, the distinct power-law scaling of the geometric distance $D$ (scaling as $h^4$ for small $h$ and $h^{-3}$ for large $h$) strongly suggests that GNME persists for all finite non-zero fields.

To provide further support for this statement, we analyze the states in the asymptotic limits $h=0$ and $h \to \infty$. We demonstrate that in both limits, the reduced density matrices lie exactly on the boundary of the set of network states $\mathcal{N}_{\text{net}}$. 
First, we note that the 6-spin RDMs at the two limits are both permutationally symmetric with respect to the three parties (where each party consists of two adjacent spins) and are fully separable. Identifying the state of each 2-spin party as a logical subsystem ($\{00,01,10,11\}$ maps to $\{0,1,2,3\}$), the limits are:
\begin{equation}
    \rho_{h=0} = \frac{1}{2}\left(\ket{000}\bra{000} + \ket{111}\bra{111}\right),
\end{equation}
corresponding to the mixture resulting from the global GHZ ground state at zero field, and
\begin{equation}
    \rho_{h\to\infty} = \ket{000}\bra{000},
\end{equation}
corresponding to the product state of the paramagnetic phase. Since fully separable states are prepared by a network with trivial resources (dimension 1), both limits belong to $\mathcal{N}_{\text{net}}$.
To prove they reside on the boundary, we utilize the theorem in Ref.~\cite{Hansenne2022}, which states that permutationally symmetric states are either fully separable or possess GNME. Consider a symmetric perturbation obtained by mixing the state with an orthogonal GHZ-type component, for instance, 
\begin{equation}
    \ket{\psi} = \frac{\ket{222}+\ket{333}}{\sqrt{2}},
\end{equation}
where $\ket{2}$ and $\ket{3}$ represent orthogonal levels in the 4-dimensional Hilbert space of each 2-spin party. Mixing this component with infinitesimal weight introduces entanglement across the bipartition $A|BC$ and symmetric cuts, which can be seen from taking a partial transpose along one of the parties.
Since the partial transpose does not mix the two sectors ${0,1}$ and ${2,3}$, the entanglement negativity of the whole state reduces to that of the orthogonal GHZ part, which is finite.
According to the theorem, since the perturbed state remains permutationally symmetric but is no longer fully separable, it must possess GNME. This implies the original states $\rho_{h=0}$ and $\rho_{h\to\infty}$ cannot be in the interior of the network set and must therefore lie on the boundary. Together, the certified GNME in the wide range $h \in [0.46, 10]$ and this boundary behaviour at the limits provide compelling evidence that the 6-spin subregion exhibits GNME for all finite transverse fields $h \neq 0$.

\subsubsection{GNME for Three qubit states}\label{sup:three_qubit}

We first describe how we estimate GNME for three-qubit states. 
Given a three-qubit state $\rho$, we construct the extended state 
$\rho_{\mathrm{ext}} = \rho \otimes \mathbb{I}/D$, 
where the ancilla is a three-qubit maximally mixed state with $D=8$. 
We then apply the six-qubit Gilbert algorithm to find the closest network state $\rho_{1,\mathrm{ext}}$ to $\rho_{\mathrm{ext}}$. 
Since the set of network states is closed under LOSR, tracing out the ancilla yields another valid network state in the three-qubit Hilbert space
$\rho_1 = \operatorname{Tr}_{\mathrm{anc}}(\rho_{1,\mathrm{ext}})$. 
The Hilbert--Schmidt distance $\mathcal{D} = \|\rho - \rho_1\|_2$ then provides an upper bound on the geometric distance to the (three-qubit) network set. 
In addition, we use the inflation technique to certify the presence of GNME and determine the corresponding noise thresholds $p_c$. 

\begin{figure}[hbt!]
    \centering
    \includegraphics[width=0.6\linewidth]{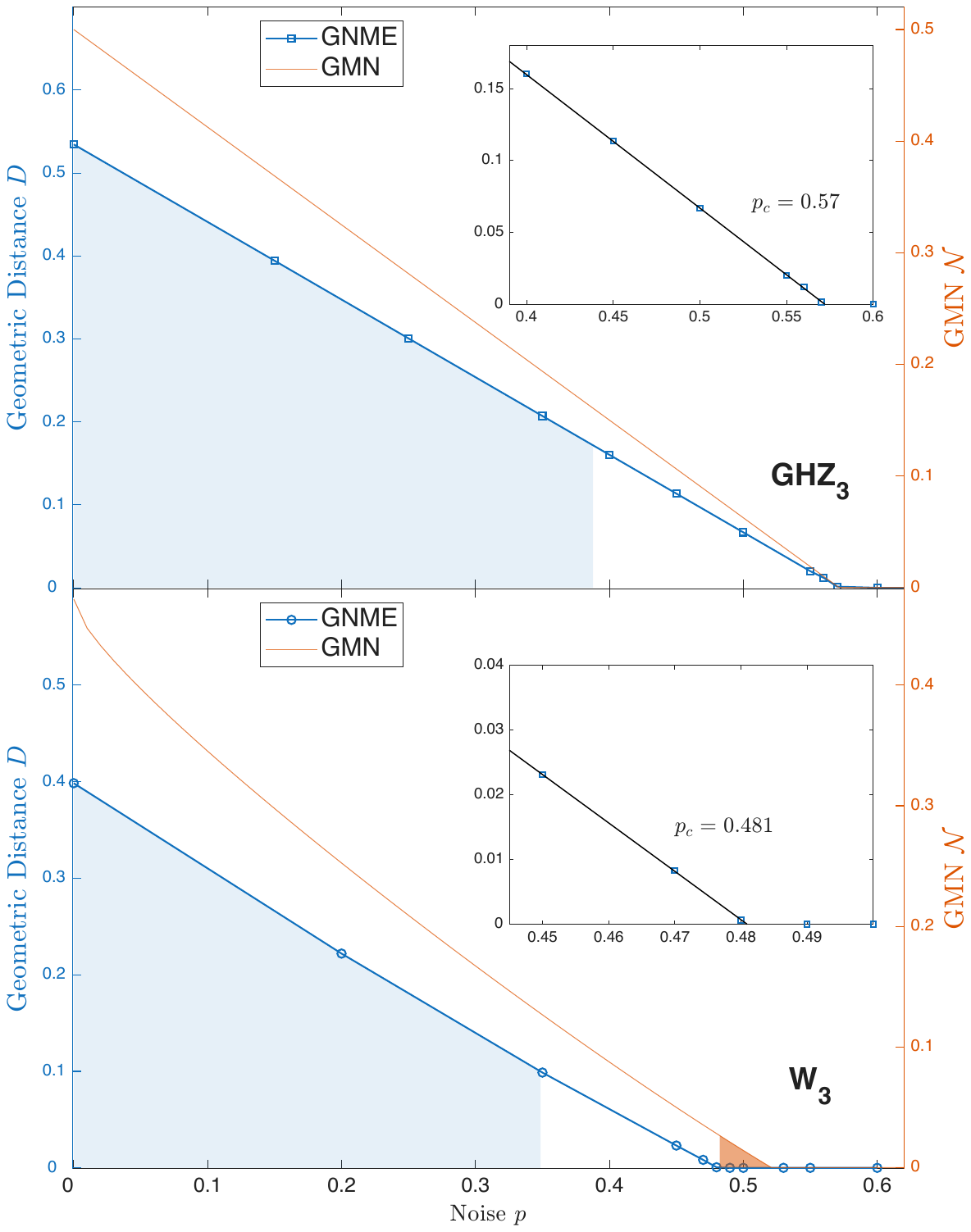}
    \caption{\textbf{GNME and GME vs white noise for three qubit states.} Shown vs the noise fraction $p$ in $\rho(p)=(1-p)\ket{\psi}\bra{\psi}+p\,\mathbb{I}/D$. 
    Upper panel: tripartite entanglement of $\ket{\text{GHZ}_3}$. Lower panel: tripartite entanglement of $\ket{W_3}$. 
    Blue curves: geometric distance to the network set (left axis); red curves: GMN (right axis). Red shaded bands mark ranges of $p$ where GME is finite, but we can certify the state to be a network state.
    Blue shaded bands refer to ranges of $p$  where GNME is certified by the inflation technique.
    }
    \label{fig:ghz_dist_3qubits}
\end{figure}

Figure~\ref{fig:ghz_dist_3qubits} illustrates the results for the mixed states 
$\rho(p) = (1-p)\ket{\psi}\bra{\psi} + p\,\mathbb{I}/D$ constructed from $\ket{\mathrm{GHZ}_3}$ (top) and $\ket{W_3}$ (bottom). 
The blue curves show the geometric distance to the network set, while the red curves show the GMN for comparison. 
Compared with the six- and eight-qubit cases discussed in the main text, the GNME and GMN curves here are closer and exhibit similar robustness thresholds, 
indicating that for three qubits, the two measures differ only slightly. 
For $\ket{\mathrm{GHZ}_3}$, the GNME threshold estimated from the geometric distance coincides with the biseparable threshold 
(for three qubits, vanishing GMN implies biseparability for permutationally invariant states~\cite{Novo2013}). 
For $\ket{W_3}$, however, we find a finite noise window $p \in [0.4824,\,0.5211]$ 
where GNME is absent while GMN remains finite. The lower bound here is given by the Gilbert criterion, discussed in Section~\ref{sup:certifyNet} of the supplemental materials.
In particular, the mixed state 
\begin{equation}
\rho = \frac{1}{2}\left(\ket{W_3}\bra{W_3} + \frac{\mathbb{I}}{8}\right)
\end{equation}
provides a simple, explicit example of a three-qubit state that is GME but not GNME. 
Such states were predicted in Ref.~\cite{navascues2020genuine} using variational methods, but no explicit examples were previously reported.

\subsubsection{Comparison with Previous Best Bounds for GNME}~\label{sup:compare_lower_bound}

We benchmark the performance of our improved inflation schemes against the best previously known bounds for GNME robustness. For three-qubit states, the strongest existing bounds were established using the original inflation framework in Ref.~\cite{navascues2020genuine}, up to second level inflation networks. From the fidelity bound, which are inequalities of rhe form $\langle\Psi\vert\rho_{net}\vert\Psi\rangle \leq F_c$ for network states, we can certify $\rho=(1-p)\vert\Psi\rangle\langle\Psi\vert+p \;(I/D)$ to be GNME for $p< p_c$ where
\begin{equation}
    p_c = \frac{1-F_c}{1-1/D}
\end{equation}
and $D$ is the dimension of the input state $\vert\Psi\rangle$. 
As shown in Table~\ref{tab:comparison}, our third-order inflation yield tighter certification windows (higher robustness thresholds $p_c$) for the $W_3$ states.

For larger systems, such as the 6-qubit states, explicit noise thresholds for GNME were not previously available in the literature. To provide a baseline for comparison, we exploit the property that GNME cannot be created by Local Operations and Shared Randomness (LOSR). By applying a local encoding map $\Lambda$ (e.g., mapping two physical qubits to one logical qubit) that transforms the 6-qubit state $\rho_6$ into the 3-qubit state $\rho_3$, any certification of GNME in the output $\rho_3$ necessarily implies GNME in the input $\rho_6$. Thus, we use the known bounds for the 3-qubit states~\cite{navascues2020genuine} as the ``previous best" lower bounds for the 6-qubit cases. Our direct 6-qubit inflation results significantly outperform these inferred bounds, demonstrating the advantage of analyzing the full system directly. Finally, for the 8-qubit Dicke state, no prior bounds exist, highlighting the capability of our method to tackle systems previously out of reach.

\begin{table}[h]
    \centering
    \begin{tabular}{l|c|c}
    \hline
    State & Inflation Lower Bound & Previous Lower Bounds \\
    \hline
    $\mathrm{GHZ}_3$ & 0.3878 &  0.4366~\cite{neumann2025} \\
        $W_3$        & 0.3486 &  0.2741~\cite{navascues2020genuine}  \\
    \hline
    $\mathrm{GHZ}_6$ & 0.5121 & $0.4366^\dagger$ \\ 
    $W_6$            & 0.4951 & $0.2741^\dagger$ \\
    \hline
    \hline
    $D_{4,8}$ & 0.2408 & -\\ 
    \hline
    \end{tabular}
    \tabnotetext{\dagger}{Certified by first mapping to their three-qubit counterparts via a local encoding map, and then applying the known three-qubit bounds.}
    \caption{\textbf{Comparison of GNME robustness lower bounds ($p_c$) obtained in this work vs. previous works.} The inflation bounds derived here (using higher-order and symmetrized inflation) consistently surpass the previous best bounds derived from Ref.~\cite{navascues2020genuine}.
    }\label{tab:comparison}
\end{table}




\subsubsection{Scaling of the geometric distance with field near $h_c = 1$}\label{sup:Ising_scaling_h}

\begin{figure*}[t]
\centering
\subfigure[]{
    \includegraphics[width=0.45\textwidth]{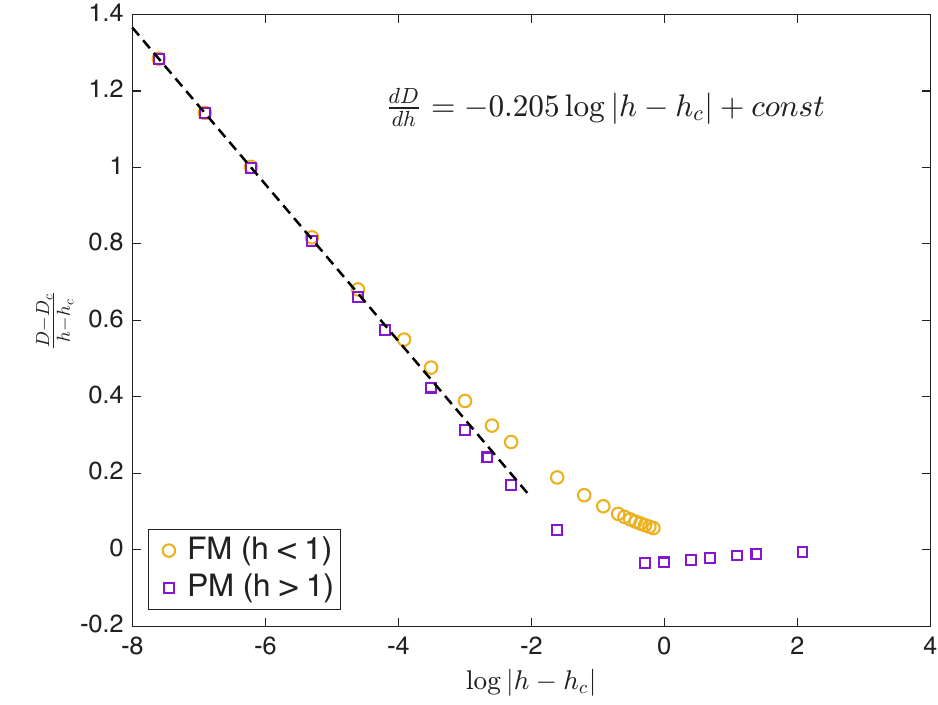}
}
\qquad
\subfigure[]{
    \includegraphics[width=0.45\textwidth]{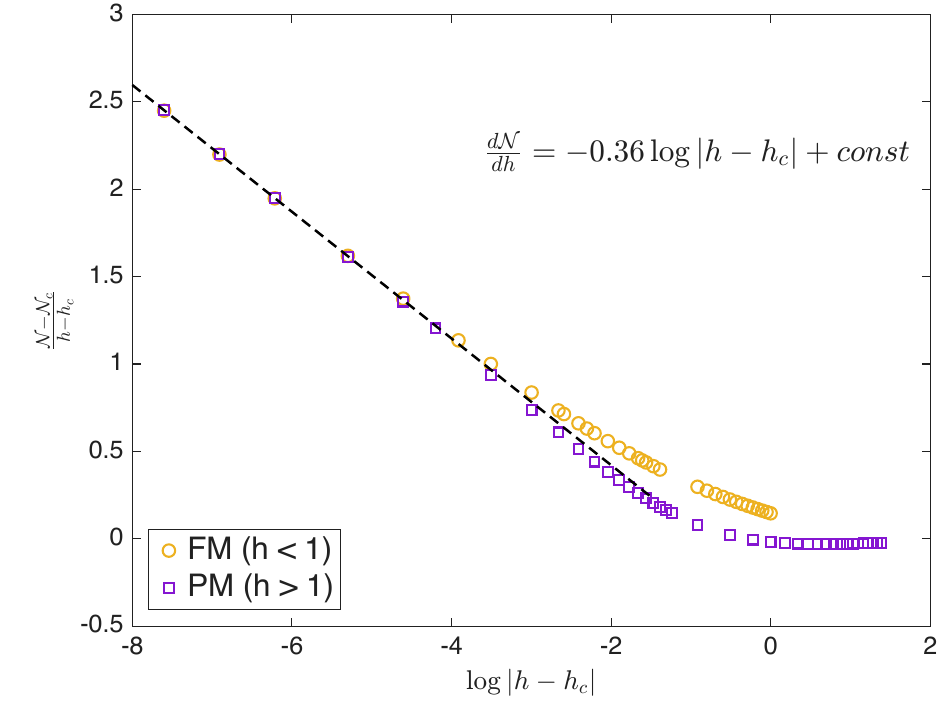}
}
\caption{\textbf{Logarithmic Divergence of the derivative for GMN and GNME near the Ising critical point.}
Plot of $(D - D_c)/(h - h_c)$ and $(\mathcal{N} - \mathcal{N_c})/(h - h_c)$ versus $\log|h - h_c|$ for the six-spin subregion.
The FM ($h<1$) and PM ($h>1$) branches are shown separately. The linear fit is obtained on the PM branch.}
\label{fig:Supplement_Ising_h_scaling}
\end{figure*}

In this subsection, we analyze how the geometric distance $D$ varies with the transverse field near the quantum critical point $h_c = 1$. We consider the transformed quantity
\[
\frac{D - D_c}{h - h_c},
\]
which approaches the derivative $dD/dh$ as $h \to h_c$. Plotting this ratio against $\log|h - h_c|$ isolates the critical behaviour and separates the FM ($h<1$) and PM ($h>1$) branches, as shown in Fig.~\ref{fig:Supplement_Ising_h_scaling}. From the linear regime of the PM branch, we extract the scaling form
\[
\frac{dD}{dh} = -0.205\,\log|h - h_c| + \text{const},
\]
which characterizes the singular slope of $D$ at the critical field.
Our result demonstrates that the geometric distance exhibits the same logarithmic critical scaling as the GMN behaviour reported in Ref.~\cite{Hofmann2014Scaling}.

\subsection{Genuine Multiparty Negativity}\label{sup:gmn}
\subsubsection{Semidefinite program for Genuine multiparty negativity}
Here, we present the semidefinite-program formulation of the genuine multiparty negativity (GMN), which we use as a GME measure, used to compare with GNME measures.
To quantify Genuine multiparty entanglement (GME), we use the Genuine Multipartite Negativity (GMN), which is an extension of the bipartite quantum negativity. First define the minimal bipartite negativity $N^{\text{min}}(\rho)= \min_m {N_m(\rho)}$ where $N_m$ is the bipartite negativity along bipartition $m$. GMN is the mixed convex roof extension of $N^{\min}$
$$
\mathcal{N}(\rho)=\min _{\left\{p_k, \rho_k\right\}} \sum_k p_k N^{\min }\left(\rho_k\right)
$$
where each ensemble $\left\{p_k, \rho_k\right\}$ satisfies $\rho=\sum_k p_k \rho_k$. GMN can be formulated via the following semidefinite program (SDP)~\cite{Guhne2011, Hofmann2014}
\begin{equation}
\begin{aligned}
\mathcal{N}(\rho)&=-\min \operatorname{tr} (\rho W) \\
\text { subject to } & W=P_m+Q_m^{T_m} \\
& 0 \leqslant P_m  \\
& 0 \leqslant Q_m \leqslant I \text {  for all bipartitions } m \mid \bar{m}
\end{aligned}    
\end{equation}
where $W$ is an entanglement witness that is fully decomposable with respect to all bipartitions. 
For two operators $A$ and $B$, the majorization $A \succeq B$ implies that $A-B$ is positive semidefinite. $T_m$ refers to a partial transpose with respect to either part of the bipartition $m |\bar{m}$. 
Importantly, since the problem is an SDP, it can be efficiently solved, and the global optimality is guaranteed. The SDP program is modeled using the YALMIP toolbox~\cite{YALMIP2004} in Matlab, and the solver used is either Mosek~\cite{mosek} (for 6 spins or less) or the splitting conic solver (SCS)~\cite{Brendan2016scs} (for more than 6 spins).

\subsubsection{Genuine multiparty Entanglement and Minimal bipartite negativity in the quantum Ising model}\label{sup:GME_ans_bipartite}
In this subsection, we compare the genuine multiparty negativity (GMN) with the minimal bipartite negativity $N_{\min}$ in the 1d quantum Ising model. 
Recall that by definition, GMN is always upper-bounded by $N_{\min}$. 
For six consecutive spins divided into three parties, $N_{\min}$ is simply the bipartite negativity across the cut $A|BC$. Thus in this simple 1d geometry, the definition of $N_{\min}$ reduces to a single bipartition. 

\begin{figure}[hbt!]
    \centering
    \begin{overpic}[width=0.5\linewidth]{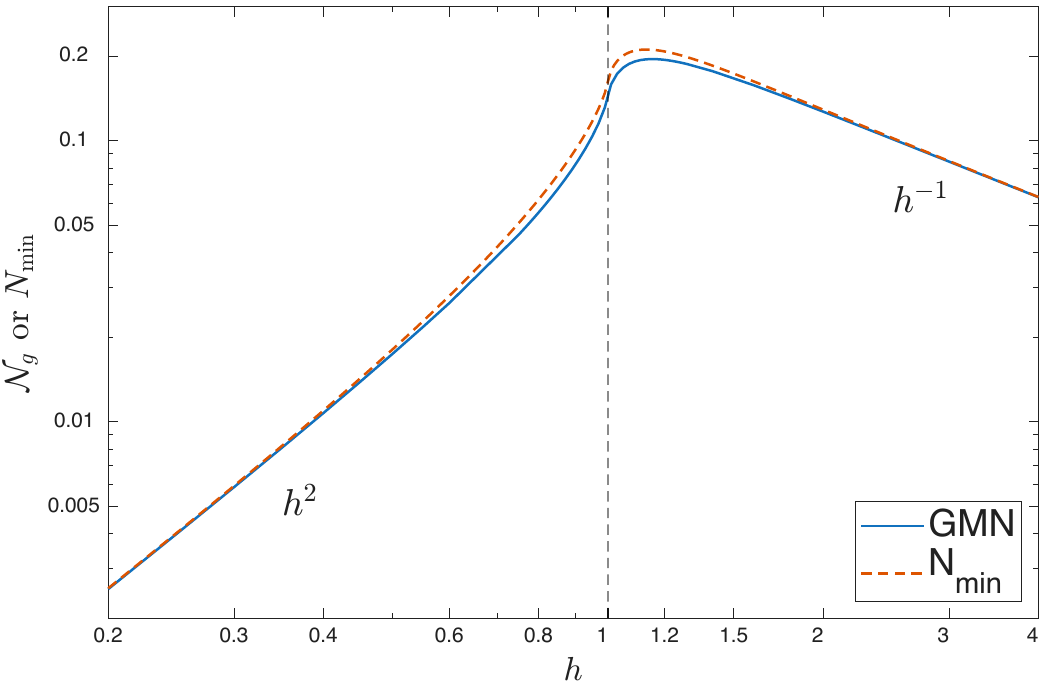}
        \put(15,50){\includegraphics[width=0.15\linewidth]{Figures/Ising_222_subregion.pdf}}
    \end{overpic}
    \caption{\textbf{Genuine multiparty negativity and minimal bipartite negativity in the 1d Quantum Ising model.} 
    GMN (blue) and minimal bipartite negativity $N_{\min}$ (red) for the 6 consecutive spins of the Ising chain. 
    The diagram illustrates the subregion partitions. 
    Both GMN and $N_{\min}$ display identical scaling across the phase transition: $h^2$ at small field and $h^{-1}$ at large field. 
    This shows that in the Ising chain, GMN essentially tracks bipartite entanglement.}
    \label{fig:Ising_neg}
\end{figure}

Figure~\ref{fig:Ising_neg} shows the comparison of GMN and $N_{\min}$. The GME measure and the bipartite entanglement measure display identical behaviour across the phase transition, collapsing exactly onto each other for small and large fields.  
On the log--log scale, we can show that they both scale as $h^2$ at small field and as $h^{-1}$ at large field. A similar scaling can also be found in the four-party entanglement of 8 consecutive spins.
This provides a clear illustration of the main point emphasized in the text: in the Ising chain, genuine multiparty entanglement is essentially dominated by bipartite contributions. 
This is in stark contrast with Figure~\ref{fig:Ising_dist} in the main text, where GMN and GNME show very different behaviours: GNME exhibits exponential scaling at small field and $h^{-3}$ scaling at large field. 
Thus, in this 1d Ising model, GNME successfully removes the trivial bipartite contribution contained in GME. 

An intuitive reason why GMN follows the bipartite scaling in this example can be understood from the alternative definition of GMN given in Ref.~\cite{Hofmann2014}: 
\begin{equation}
    N_g(\varrho)=\min _{\varrho=\sum_m p_m \varrho_m} \sum_m p_m N_m\left(\varrho_m\right),
\end{equation}
where $m$ runs over all inequivalent bipartitions of the $n$ parties, giving $2^{n-1}-1$ possible bipartitions. 
From this perspective, GMN seeks a decomposition of the state that minimizes the average bipartite negativity. 
For certain geometries, such as the 1d subregions studied here, there is a clear preferred bipartition where the negativity is smallest—namely the cut that isolates the boundary party $A$ from the rest. 
As a result, the optimal decomposition places most of its weight on this bipartition, and the GMN reduces to the negativity across $A|\text{rest}$. 
This explains why, in Fig.~\ref{fig:Ising_neg}, GMN and $N_{\min}$ coincide and both scale as $h^2$ at small field and $h^{-1}$ at large field. 

Technically, the bipartite negativity $\mathcal{N}$ is not additive, while the \emph{logarithmic negativity} $E_{\mathcal{N}}=\log(2\mathcal{N}+1)$ is~\cite{Vidal2002LogNeg}. Consequently, it is $E_{\mathcal{N}}$ that rigorously satisfies an area law, and by extension, the \emph{logarithmic genuine multiparty negativity} $\log(2N_g+1)$ should exhibit area-law scaling. Throughout this work, however, we use the standard (non-logarithmic) GMN, as it remains the conventional quantity in the quantum information literature. This choice is justified here because the subregions of many-body systems we consider contain only a few spins, and have a magnitude of GMN $N_g<1$, so the behaviours of $N_g$ and $\log(2N_g+1)$ are quantitatively similar.

\begin{figure}[t]
\centering
\includegraphics[width=0.5\linewidth]{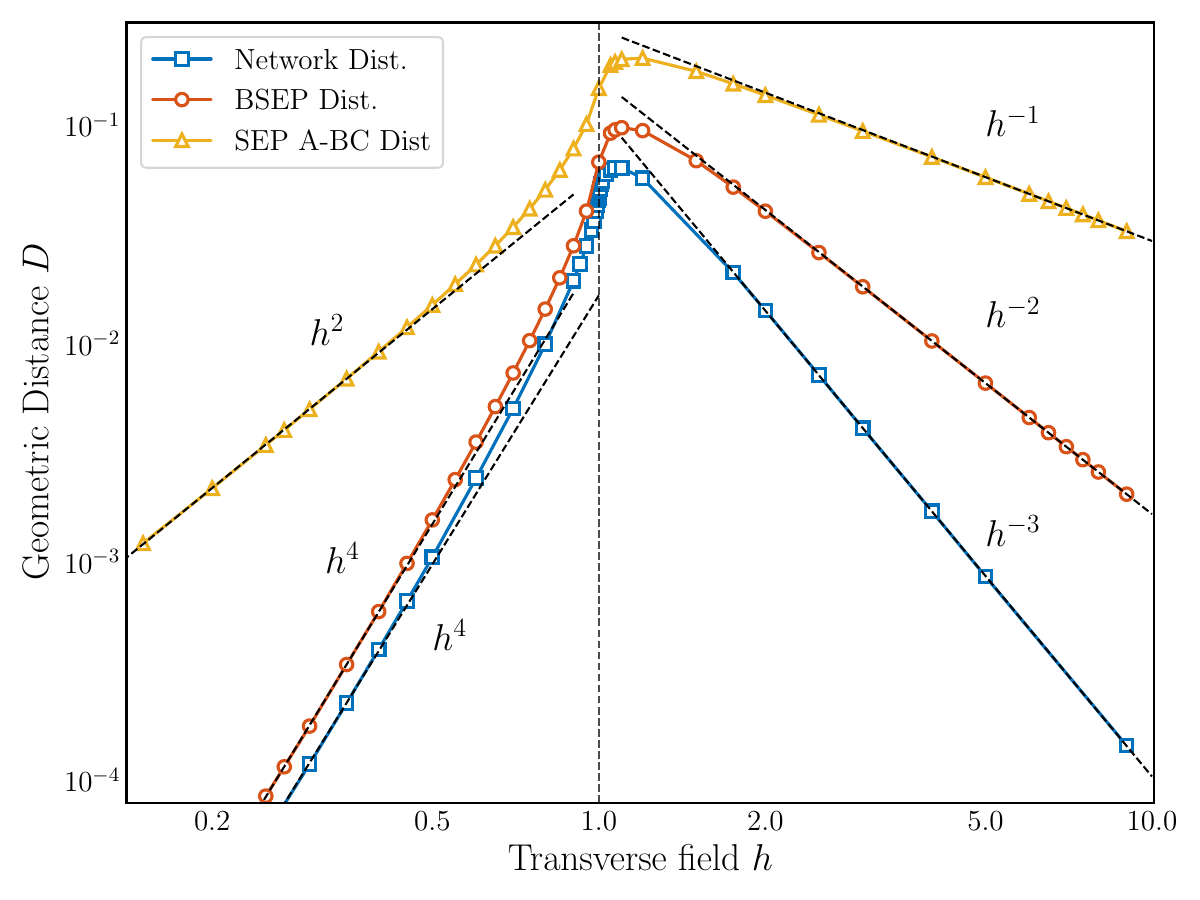}
\caption{\textbf{Comparison of geometric distances $D$ in the 1D transverse field Ising model.} We plot the minimal Hilbert-Schmidt distances from the 6-spin
reduced density matrix to the set of bipartite separable states (2|4 partition), biseparable states, and the unitary quantum network set. Dashed lines indicate power-law fits $D \propto h^k$. The SEP distance scales as $h^2$ and $h^{-1}$ respectively. The BSE distance (representing GME) exhibits higher scaling exponents ($h^4$ and $h^{-2}$), while the Network distance (representing GNME) matches the $h^4$ scaling at small field but exhibits a sharper decay ($h^{-3}$) in the paramagnetic phase ($h \gg 1$).}
\label{fig:GME_GNME_comparison}
\end{figure}
To provide a direct comparison with the network set across a unified metric, we evaluate the geometric distance $D$ (Fig.~\ref{fig:GME_GNME_comparison}) to the bipartite separable (SEP) and biseparable (BSEP) sets using the same Gilbert algorithm and compare them with the distance to the network set. We find that the distance to the bipartite separable (SEP) set scales as $h^2$ and $h^{-1}$, perfectly mirroring the behavior of the bipartite negativity and GMN analyzed previously. Interestingly, the distance to the biseparable set (representing GME) exhibits steeper scaling exponents ($h^4$ and $h^{-2}$) than the GMN, highlighting that different GME measures can yield distinct scaling powers. Furthermore, this steeper scaling matches the GNME behavior at small fields. However, in the paramagnetic phase, the GNME distance decays significantly faster ($h^{-3}$) than the GME distance ($h^{-2}$). In summary, GNME still proves to be a sharper detector of criticality.  

\subsection{Scaling of 6-spin GMN with temperature}

\begin{figure}[t]
    \centering
    \includegraphics[width=0.45\linewidth]{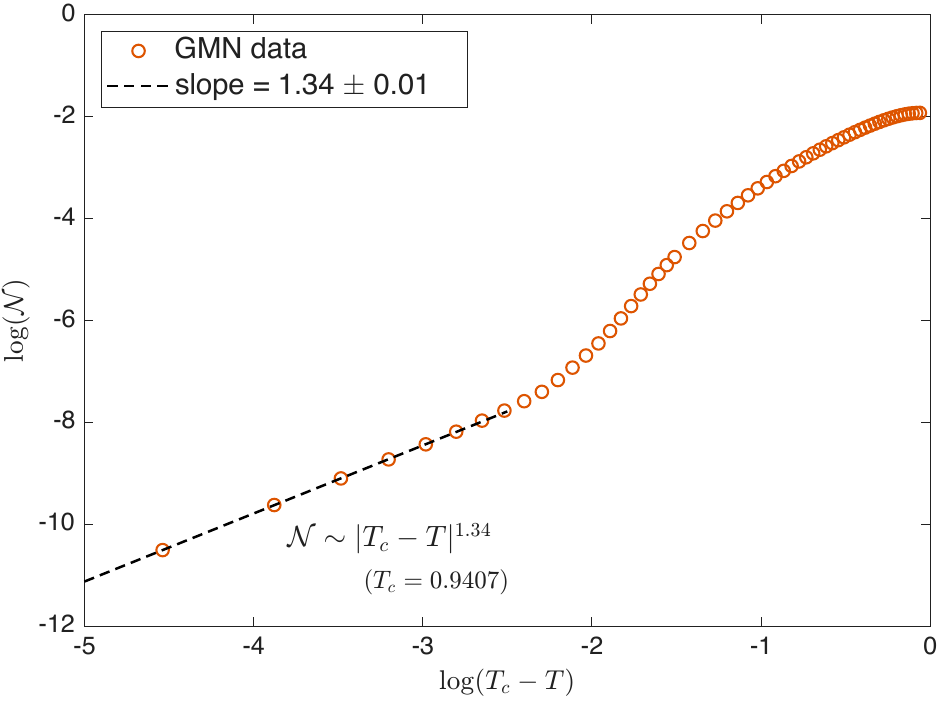}
    \caption{
        Log--log plot of the GMN for $h=1$ as a function of $T_c - T$. 
        The linear regime close to $T_c$ reveals a power-law decay 
        $\mathcal{N}(T) \propto (T_c - T)^{\alpha_{\mathrm{GMN}}}$, with a slope
        consistent with the fit used for the GMN curve in Fig.~\ref{fig:Ising_finiteT}.
    }
    \label{fig:GMN_loglog}
\end{figure}

In Fig.~\ref{fig:Ising_finiteT}, the GMN curve for $h=1$ rapidly approaches zero as $T$ approaches the thermal critical point $T_c$, which compresses its power-law decay on the linear vertical axis. To resolve this behaviour, we plot the GMN data in log--log scale as $\log \mathcal{N}$ versus $\log(T_c - T)$ in Fig.~\ref{fig:GMN_loglog}.

A clear linear regime emerges near criticality, spanning several decades in $\mathcal{N}$. Fitting this interval yields a slope $\alpha_{\mathrm{GMN}}$ that matches the value used for the GMN fit in Fig.~\ref{fig:Ising_finiteT}. This confirms that GMN follows a critical power law of the form
\[
\mathcal{N}(T) \propto (T_c - T)^{\alpha_{\mathrm{GMN}}},
\]
consistent with the scaling extracted from the GNME analysis. The supplemental figure provides a transparent visualization of this scaling that is difficult to discern in the main panel.

\subsection{Properties of mixed convex roof extensions}\label{sup:mixedCRE}

In this section, we summarize the key properties of mixed convex-roof extensions of minimal bipartite entanglement monotones and provide proofs justifying their use as measures of genuine multiparty entanglement.
Let $E_{A \mid B}$ be a bipartite entanglement monotone that satisfies three standard conditions:
\begin{enumerate}
    \item it is \textit{faithful} ($E\geq0$, zero if and only if the state is separable),
    \item it is invariant under local unitary (LU) transformations, and
    \item it is non-increasing on average under local operations and classical communication (LOCC).
\end{enumerate}
We define the minimal bipartite entanglement of a state $\rho$ as
\begin{equation}
    E_{\text{min}}(\rho)=\min_m E_m(\rho),
\end{equation}
where $m$ runs over all possible bipartitions of the system. Since each $E_m$ satisfies the properties above, $E_{\text{min}}$ is also faithful on biseparable states, LU-invariant, and LOCC monotonic.
The GME measure is defined via the mixed convex roof extension:
\begin{equation}
    \mathcal{E}(\rho)=\min _{\{p_k,\rho_k\}} \sum_k p_k E_{\text{min}}\left(\rho_k\right),
    \label{eq:CRE_GME}
\end{equation}
where the minimization is taken over all possible decompositions of the state $\rho=\sum_k p_k \rho_k$. We now explicitly prove the properties of $\mathcal{E}(\rho)$.

\begin{enumerate}
    \item \textbf{Faithfulness ($\mathcal{E}(\rho)=0$ iff $\rho$ is biseparable).} \\
    If $\rho$ is biseparable, there exists a decomposition $\rho = \sum_k p_k \rho_k$ such that each $\rho_k$ is separable with respect to some specific bipartition $m_k$. Consequently, $E_{m_k}(\rho_k) = 0$ for each $k$. Since $E_{\text{min}}(\rho_k) \le E_{m_k}(\rho_k)$ and $E \ge 0$, it follows that $E_{\text{min}}(\rho_k) = 0$ for all $k$. Therefore, the convex roof $\mathcal{E}(\rho)$ vanishes.
    
    Conversely, assume $\mathcal{E}(\rho) = 0$. Since $E_{\text{min}} \ge 0$, the optimal decomposition $\{p_k, \rho_k\}$ must satisfy $E_{\text{min}}(\rho_k) = 0$ for all $k$. This implies that for every $\rho_k$, there exists at least one bipartition $m_k$ such that $E_{m_k}(\rho_k) = 0$. Because the underlying bipartite measure is faithful, $\rho_k$ must be separable with respect to the cut $m_k$. Thus, every state $\rho_k$ in the decomposition is biseparable. Since $\rho$ is a convex mixture of biseparable states, $\rho$ is itself biseparable.

    \item \textbf{Invariance under Local Unitaries.} \\
    Let $\rho' = U_{\text{loc}} \rho U_{\text{loc}}^\dagger$. Suppose $\{p_k, \rho_k\}$ is the optimal decomposition for $\rho$ such that $\mathcal{E}(\rho) = \sum_k p_k E_{\text{min}}(\rho_k)$. The set $\{p_k, \rho'_k = U_{\text{loc}} \rho_k U_{\text{loc}}^\dagger\}$ forms a valid decomposition for $\rho'$. Due to the LU-invariance of the underlying bipartite measure, $E_{\text{min}}(\rho'_k) = E_{\text{min}}(\rho_k)$. Since $\mathcal{E}(\rho')$ is minimized over all decompositions, we have $\mathcal{E}(\rho') \le \mathcal{E}(\rho)$. By symmetry, applying $U_{\text{loc}}^\dagger$ to $\rho'$ recovers $\rho$, implying $\mathcal{E}(\rho) \le \mathcal{E}(\rho')$. Thus, $\mathcal{E}(\rho) = \mathcal{E}(\rho')$.

    \item \textbf{LOCC Monotonicity.} \\
    Let $\Lambda$ be an LOCC map. Suppose $\{p_k, \rho_k\}$ is the optimal decomposition for $\rho$. By linearity, $\Lambda(\rho) = \sum_k p_k \Lambda(\rho_k)$ is a valid decomposition of the output state. By definition, $\mathcal{E}(\Lambda(\rho))$ is upper bounded by the value of this specific decomposition:
    \begin{equation}
        \mathcal{E}(\Lambda(\rho)) \le \sum_k p_k E_{\text{min}}(\Lambda(\rho_k)).
    \end{equation}
    Since the underlying bipartite measures are non-increasing under LOCC, $E_{\text{min}}(\Lambda(\rho_k)) \le E_{\text{min}}(\rho_k)$. It follows immediately that $\mathcal{E}(\Lambda(\rho)) \le \sum_k p_k E_{\text{min}}(\rho_k) = \mathcal{E}(\rho)$.

    \item \textbf{Convexity.} \\
    Let $\rho = \lambda \sigma + (1-\lambda) \tau$. Let $\{q_i, \sigma_i\}$ be the optimal decomposition for $\sigma$ and $\{r_j, \tau_j\}$ be the optimal decomposition for $\tau$. We can construct a valid decomposition for $\rho$ by combining these ensembles with weights $\lambda q_i$ and $(1-\lambda)r_j$. The value of $\mathcal{E}(\rho)$ is bounded by this specific decomposition:
    \begin{align}
        \mathcal{E}(\rho) &\le \sum_i \lambda q_i E_{\text{min}}(\sigma_i) + \sum_j (1-\lambda) r_j E_{\text{min}}(\tau_j) \\
        &= \lambda \mathcal{E}(\sigma) + (1-\lambda) \mathcal{E}(\tau).
    \end{align}
    Thus, $\mathcal{E}$ is convex.
\end{enumerate}

\subsection{Properties of the network geometric entanglement}\label{sup:properties_GNME_dist}

In this section, we discuss the theoretical properties of the geometric distance to the set of network states. Let $d(\rho, \sigma)$ be a generic distance metric induced by a matrix norm. The geometric network entanglement of a state $\rho$ is defined as the minimum distance to the set of network states:

\begin{equation}
    \mathcal{D}_{\text{net}} = \min_{\sigma \in \text{net}} d(\rho, \sigma)
\end{equation}
Depending on the choice of the norm, this geometric distance possesses different information-theoretic properties. In what follows, we specifically contrast the distance induced by the trace norm, $d_{tr}(\rho, \sigma) = \|\rho - \sigma\|_{tr}=tr\sqrt{(\rho-\sigma)^\dagger (\rho-\sigma)}$, with the Hilbert-Schmidt (HS) distance used in our Gilbart algorithm, $d_{HS}(\rho, \sigma) = \|\rho - \sigma\|_{HS}=\sqrt{tr(\rho-\sigma)^\dagger (\rho-\sigma)}$.

We consider several key properties of a GNME measure. For the geometric distance $\mathcal{D}_{\text{net}}$, these evaluate as follows:

\begin{itemize}
\item \textbf{Faithfulness:} Because the set of network states is topologically closed, $\mathcal{D}_{\text{net}} = 0$ if and only if $\rho$ is a network state.
\item \textbf{Convexity:} The set of network states is convex due to shared randomness. Since both the trace and HS distances are induced by valid matrix norms, the resulting minimum distance $\mathcal{D}_{\text{net}}$ is strictly convex.
\item \textbf{Monotonicity under LOSR:} To be a valid entanglement monotone, the underlying metric must be contractive under LOSR. Monotonicity is true if we use the trace norm but not the HS norm, as explained below.
\end{itemize}

If the geometric distance is defined using the trace norm, since the trace distance is contractive under any CPTP map \cite{RUSKAI_tracenorm_contractivity},
\begin{equation}
    d_{\mathrm{tr}}\left(\mathcal{E}\left(\rho_1\right), \mathcal{E}\left(\rho_2\right)\right) \leq d_{\mathrm{tr}}\left(\rho_1, \rho_2\right)
\end{equation}
for any quantum channel $\mathcal{E}$. However, the trace norm is non-smooth; its gradient exhibits singularities whenever the difference matrix $\rho - \sigma$ is not full-rank. This non-differentiability causes gradient-based optimization algorithms, such as the Gilbert method we employ to estimate the distance, to fail or exhibit severe numerical instabilities. This computational bottleneck dictates our use of the HS norm, which provides the necessary smoothness for the Gilbert algorithm. While the HS distance is not strictly contractive under CPTP maps (and thus not a strict LOSR monotone), standard matrix norm inequalities guarantee that it strictly lower-bounds the trace norm:
\begin{equation}
    \|\rho - \sigma\|_{HS} \leq \|\rho - \sigma\|_{tr}    
\end{equation}
Consequently, any strictly positive HS geometric distance, $\mathcal{D}_{\text{net}} > 0$, rigorously guarantees a positive trace distance. Therefore, our computed quantity acts as a mathematically rigorous geometric entanglement witness for network entanglement.


\end{document}